\documentclass{ieeeaccess}
\usepackage{cite}
\usepackage{amsmath,amssymb,amsfonts}
\usepackage{algorithm}
\usepackage{algorithmic}
\usepackage{graphicx}
\usepackage{textcomp}
\usepackage{xcolor}
\usepackage{mathtools}
\usepackage{color,soul}
\usepackage{graphicx}
\usepackage{booktabs}
\usepackage[font={small}]{caption}
\usepackage{subcaption}
\usepackage{cite}
\usepackage{hyperref}
\usepackage{mathrsfs}
\usepackage{accents}
\usepackage{acronym}
\usepackage{bm}
\usepackage{wrapfig}

\renewcommand{\thetable}{\arabic{table}}

\DeclareMathOperator*{\argmax}{arg\,max}

\graphicspath{{./figures/}}
\newcommand{\eq}[1]{Eq.~\eqref{#1}}
\newcommand{\fig}[1]{Fig.~\ref{#1}}
\newcommand{\tab}[1]{Tab.~\ref{#1}}
\newcommand{\secref}[1]{Section~\ref{#1}}
\newcommand{\alg}[1]{Alg.~\ref{#1}}

\newcommand{\mytexttilde}{{\raise.17ex\hbox{$\scriptstyle\mathtt{\sim}$}}}

\def\BibTeX{{\rm B\kern-.05em{\sc i\kern-.025em b}\kern-.08em
    T\kern-.1667em\lower.7ex\hbox{E}\kern-.125emX}}

\begin{document}
\history{{\small \textcopyright 2021 IEEE. This work has been accepted for publication in the IEEE Access. }}
\doi{}

\title{Real-time People Tracking and Identification from Sparse mm-Wave Radar Point-clouds}
\author{
\uppercase{Jacopo Pegoraro}\authorrefmark{1}, \IEEEmembership{Graduate Student  Member, IEEE}, \uppercase{and Michele Rossi}\authorrefmark{1,2}, \IEEEmembership{Senior Member, IEEE}}
\address[1]{Department of Information Engineering, University of Padova, 35131 Padova, IT.}
\address[2]{Department of Mathematics ``Tullio Levi-Civita'', University of Padova, 35131 Padova, IT.}
\tfootnote{This work has been supported, in part, by MIUR (Italian Ministry of Education, University and Research) through the initiative "Departments of Excellence" (Law 232/2016).}

\markboth
{Author \headeretal: Preparation of Papers for IEEE TRANSACTIONS and JOURNALS}
{Author \headeretal: Preparation of Papers for IEEE TRANSACTIONS and JOURNALS}

\corresp{Corresponding author: Jacopo Pegoraro (e-mail: pegoraroja@dei.unipd.it).}

\begin{abstract}
\mbox{Mm-wave} radars have recently gathered significant attention as a means to track human movement and identify subjects from their gait characteristics. A widely adopted method to perform the identification is the extraction of the \textit{micro-Doppler signature} of the targets, which is computationally demanding in case of \mbox{co-existing} multiple targets within the monitored physical space. Such computational complexity is the main problem of \mbox{state-of-the-art} approaches, and makes them inapt for \mbox{real-time} use. In this work, we present an \mbox{end-to-end}, \mbox{low-complexity} but highly accurate method to track and identify multiple subjects in \mbox{real-time} using the sparse \mbox{point-cloud} sequences obtained from a \mbox{low-cost} \mbox{mm-wave} radar. Our proposed system features an extended object tracking Kalman filter, used to estimate the position, shape and extension of the subjects, which is integrated with a novel deep learning classifier, specifically tailored for effective feature extraction and \mbox{fast} inference on radar point-clouds. The proposed method is thoroughly evaluated on an edge-computing platform from NVIDIA (Jetson series), obtaining greatly reduced execution times (reduced complexity) against the best approaches from the literature. Specifically, it achieves accuracies as high as $91.62$\%, operating at $15$ frames per seconds, in identifying three subjects that concurrently and freely move in an unseen indoor environment, among a group of eight.
\end{abstract}

\begin{keywords}
mm-wave radar, person identification, point-clouds, multi-target tracking, convolutional neural networks
\end{keywords}

\titlepgskip=-15pt

\maketitle

\section{Introduction}
\label{sec:introduction}
The use of \mbox{mm-wave} radars for physical environment sensing is a fast growing research area~\cite{shah2019rf, patole2017automotive}.
They can be used to infer the position of the surrounding obstacles and humans, with high precision, by transmitting an interrogation signal and analyzing the modifications in the received reflected waves~\cite{knudde2017indoor}. These systems represent an effective means to monitor indoor environments, inferring key information about the movement of people, without capturing any visual image of the scene, which could rise privacy concerns. In addition, in contrast with camera surveillance systems, radars are insensitive to poor light conditions, to the presence of smoke~\cite{lu2020see}, and are also energy efficient and \mbox{low-cost} as compared to other technologies such as LIDARs~\cite{shah2019rf}. 

The high sensitivity of \mbox{mm-waves} to the frequency shifts caused by the Doppler effect makes them suitable to infer the movement patterns of humans. A widely adopted method of analysis is to extract features from the \mbox{so-called} \textit{micro-Doppler signature} ($\mu$D) of the subject, which contains \mbox{time-frequency} information about the induced Doppler shift, including the contribution of \mbox{small-scale} movements~\cite{chen2006micro,chen2000analysis}. $\mu$D signatures have been used in the challenging task of distinguishing subjects from their way of walking (\textit{gait}), which is the aim of the present work.

Human gait has been classified as a \textit{soft biometric} \mbox{\cite{nambiar2019gait}}, meaning it is unique for each person. Differently from \textit{hard biometrics}, however, such as fingerprints or DNA, it can not be used in high-stakes settings or to uniquely identify subjects among very large groups, e.g., more than $1,000$ people. Despite this, gait is difficult to fake, and it can be effectively analyzed even at distance and without requiring the subjects to collaborate. For these reasons, mm-wave radar based gait recognition can be a good option to identify subjects in scenarios such as surveillance systems or individually-tailored smart home applications, where the number of people involved is in the order of a few tens, replacing or augmenting traditional camera systems.

Using a \textit{deep learning} (DL) classifier on raw micro-Doppler spectrograms has proven to be robust and effective for identifying subjects in the case of single- and \mbox{multi-target} scenarios~\cite{vandersmissen2018indoor,yang2019person,chen2018personnel,pegoraro2020multiperson}. The extraction of representative features from $\mu$D signatures is often accomplished via neural networks (NN) and DL methods. These, are becoming the tools of choice due to the high randomness of \mbox{mm-wave} propagation, which makes a full mathematical modelling of the involved dynamics very difficult~\cite{seifert2019toward,seyfiouglu2018deep,jin2019multiple}.


In the present work, we design and validate a realtime \mbox{\it multi-target} tracking and identification system running on constrained edge-computing devices\footnote{As an example, see the NVIDIA Jetson series.} equipped with hardware accelerators (last generation GPUs). Instead of working on the raw data obtained from the backscattered \mbox{mm-wave} signal, as commonly done in the literature, we use {\it sparse point-clouds}. This makes it possible to implement our system on resource limited edge-computing devices. \mbox{Point-clouds} carry information about the \mbox{three-dimensional} spatial coordinates of the reflecting points, their velocity and the reflected power, and are obtained by employing detection algorithms at the radar processing unit, thus avoiding the need for transferring the {\it full raw data} from the radar to the edge computer. Due to their much lower data size, they bring advantages in terms of communication and computation at the connected processing device. Nonetheless, these advantages entail a {\it more challenging person identification task}: the sparsity of radar \mbox{point-cloud} data can be a source of inaccuracy and {\it standard DL architectures are inapt for learning from them}, as they rely on the reciprocal ordering of their input elements~\cite{qi2017pointnet}. As a solution, we present a novel DL classifier, called temporal convolution \mbox{point-cloud} network (TCPCN), which allows extracting meaningful order-invariant features from sparse \mbox{point-cloud} data.

The proposed system sequentially performs person tracking and identification, estimating the positions and the identities of humans as they freely move in an indoor space. For that, we use a \mbox{low-cost} Texas Instruments IWR1843BOOST \mbox{mm-wave}, \mbox{frequency-modulated} \mbox{continuous-wave} (FMCW), \mbox{multiple-input} \mbox{multiple-output} (MIMO) radar and implement the required processing functions in \mbox{real-time} on a commercial \mbox{edge-computing} node (NVIDIA Jetson series). To carry out the person identification task, we combine standard tracking techniques, i.e., Kalman filter, with DL methods. This combined use of filtering and DL makes it possible to effectively capture the time evolution of the \mbox{point-cloud} representing each subject. Our main contributions are:
\begin{enumerate}
	\item We build an end-to-end tracking and identification system that reliably operates in \mbox{real-time} at over $15$~fps on a commercial edge-computing device paired with a low-cost mm-wave radar. The approach reaches an accuracy of $91.62$\% in identifying up to three subjects (among a group of eight) freely and concurrently moving in a new indoor space, i.e., not seen at training time.
	\item We propose a novel DL classifier, called \textit{temporal convolution point-cloud network} (TCPCN), that is tailored on mm-wave radar point-cloud sequences and that is both accurate and fast. TCPCN contains a feature extraction block that obtains global information from the radar output at each time-frame and a block that exploits causal dilated convolutions~\cite{oord2016wavenet} to recognize meaningful patterns in the temporal evolution of the features. Our model significantly outperforms state-of-the-art neural networks in this field in terms of classification accuracy and inference time.
	\item The tracking phase of our system employs a \mbox{converted-measurements} Kalman filter (CM-KF) that, in addition to estimating the position of the targets in Cartesian coordinates, also estimates the extension of the subject in the horizontal plane ($x-y$), considering him/her as an {\it extended object} rather than an ideal \mbox{point-shaped} reflector. This provides useful additional information that could be exploited by, e.g., occupancy or proximity based applications. In fact, knowing the extension of the subjects would be valuable for \textit{(i)} smart-home applications that perform occupancy detection in certain areas, \textit{(ii)} security systems in industrial settings, to estimate how close a person is to some dangerous area or machinery, \textit{(iii)} detection systems (e.g., for automatic gates) that could quickly discern between cars, adults, kids or pets from their size. To the best of our knowledge, no earlier work uses \textit{extended object tracking} (EOT) within a \mbox{point-cloud} based tracking and identification system.
\end{enumerate}

The novelty of the proposed solution stems from the following main points: the design and implementation of a novel \mbox{DL-based} neural network classifier working on time sequences of sparse \mbox{point-cloud} data, that is at the same time {\it highly accurate} and {\it fast}, the integration of tracking and identification phases, that in the literature on the subject are usually dealt with separately, the implementation and validation of the solution on a commercially available \mbox{edge-computing} platform with limited capacity.

The rest of the paper is structured as follows. In \secref{sec:rel-work}, the literature on person identification using mm-wave radars is reviewed, underlining the novel aspects in our approach. In \secref{sec:mmwave-model}, the FMCW MIMO radar signal model is outlined, by also describing the procedure to extract the point-clouds. Our proposed framework is presented in \secref{sec:system}. In \secref{sec:results}, experimental results are shown, while concluding remarks are given in \secref{sec:conclusion}. 

%

\section{Related Work}
\label{sec:rel-work}

In the last few years, person identification from backscattered \mbox{mm-wave} radio signals has attracted a considerable and growing interest.
Most of the research attention has been paid to processing human micro-Doppler ($\mu$D) signatures as a means to distinguish among subjects, usually employing deep learning classifiers, applied to the $\mu$D spectrogram~\cite{vandersmissen2018indoor, cao2018radar,yang2019person,abdulatif2019person,chen2018personnel,jalalvand2019radar,polfliet2018structured,pegoraro2020multiperson}.  
Although this approach is robust and accurate, it presents some drawbacks. First, the extraction of $\mu$D signatures in case of multiple targets is a rather complex endeavour, and most of the above referenced solutions only work for a \mbox{single-subject}. In very few works, e.g.,~\cite{pegoraro2020multiperson}, the authors devised methods to \mbox{single-out} the contribution from multiple concurrent targets, obtaining the individual $\mu$D signatures. However, in the interest of obtaining highly accurate signatures, these previous algorithms dealt with \mbox{\it non-sparse} radar range-azimuth-Doppler (RDA) maps that require a large communication bandwidth to transfer the raw radio data from the radar to the processing device, preventing their implementation on \mbox{low-cost} embedded boards.

Only a few works so far have considered \mbox{point-clouds} obtained from a low-cost mm-wave MIMO radar device. The sparsity of radar point-cloud data makes the identification task more challenging, as the specific features that identify each subject are more difficult to extract, and more sensitive to external disturbances. In~\cite{zhao2019mid}, a recurrent neural network with long short-term memory (LSTM) cells is used for the identification. The overall accuracy obtained for $12$ subjects is around $89$\%, and evidence that the system is able to distinguish between two concurrently walking subjects is provided. However, no evaluation of the accuracy is conducted when more than $2$ subjects share the same physical space, nor by testing it in a different indoor environment (e.g., a new room) after its training. In addition, the \mbox{point-cloud} nature of the radar data is not fully exploited: the velocity and the received power are not used, and the classifier network requires the input data to be mapped onto a $3$D voxel representation, which is inefficient and computationally expensive. The authors of~\cite{meng2020gait} proposed a deep learning model that outperforms the \mbox{bi-directional} LSTM in~\cite{zhao2019mid} on their dataset. Two radar devices are used, transmitting and receiving simultaneously, leading to an increased field of view in case of blockage. However, robust methods are neither provided for tracking multiple subjects, e.g., Kalman or particle filtering \cite{kalman1960new}, nor to reliably associate the detections (user identities) with trajectories. This seriously impacts the identification performance when multiple targets freely move in the monitored environment. In~\cite{meng2020gait}, it is in fact reported that the accuracy drops to $45$\% in a \mbox{multi-target} setting.

With the present work, we fill a literature gap, by designing a system that performs accurate tracking of multiple subjects from their point-clouds. Extended object tracking based on Kalman filtering is exploited in conjunction with a fast and novel \mbox{domain-specific} deep learning classifier. A tight integration of the tracking and identification modules is sought, towards enhancing the identification robustness and avoiding wrong identity associations and trajectory swaps. Moreover, and to the best of our knowledge, we are the first to provide an empirical study on the feasibility of operating the system in realtime on commercial \mbox{edge-computing} devices, and \mbox{low-cost} \mbox{mm-wave} radars.

\section{mmWave Radar Signal Processing}\label{sec:mmwave-model}

A \textit{frequency-modulated continuous wave} (FMCW) radar allows the joint estimation of the distance and the radial velocity of the target with respect to the radar device.
This is achieved by transmitting sequences of linear \emph{chirps}, i.e., sinusoidal waves with frequency that is linearly increased over time, and measuring the frequency shift of the reflected signal at the receiver. The frequency of the transmitted chirp signal is increased from a base value $f_o$ to a maximum $f_1$ in $T$ seconds. Defining the bandwidth of the chirp as \mbox{$B = f_1 - f_o$}, bandwidth $B$ and chirp duration $T$ are related through \mbox{$\zeta = B/T$}, and the transmitted signal is expressed as  
\begin{equation}
	s(t) = \exp{\left[ j 2\pi \left(f_o + \frac{\zeta}{2} t \right)t \right]}, \quad 0 \leq t \leq T.
\end{equation}
The chirps are transmitted every $T_{\mathrm{rep}}$ seconds in sequences of $L$ chirps each, so that the total duration of a transmitted (TX) sequence is $LT_{\mathrm{rep}}$. A full sequence, termed \textit{radar frame}, is repeated with period $\Delta t$. At the receiver, a mixer combines the received signal (RX) with the one transmitted, generating the intermediate frequency (IF) signal, i.e., a sinusoid whose instantaneous frequency corresponds to the difference between those of the TX and RX signals.
Each chirp is sampled with sampling period $T_f$ (referred to as \emph{fast time} sampling) obtaining $M$ points, while $L$ samples, one per chirp from adjacent chirps, are taken with period $T_{\mathrm{rep}}$  (\emph{slow time} sampling).

The use of multiple-input multiple-output (MIMO) radar devices allows the additional estimation of the angle-of-arrival (AoA) of the reflections, by computing the phase shifts between the receiver antenna elements due to their different positions (i.e., their different distances from the target). This is referred to as \emph{spatial} sampling, and enables the localization of the targets in the physical space.
The radar device used in this work has $N_{\mathrm{TX}}=3$ transmitter and $N_{\mathrm{RX}}=4$ receiver antennas, that are equivalent to a virtual receiver array of $N_{\mathrm{TX}}N_{\mathrm{RX}}=12$ antennas. The transmitting elements are arranged along two spatial dimensions, which we refer to as azimuth (AZ) and elevation (EL), and are used to transmit the chirp sequences according to a \mbox{time-division} multiplexing (TDM) scheme. This enables the estimation of the EL and AZ angles of the reflecting points. In \secref{sec:rdmap}, we first consider one of the receiver elements, referring to it as \emph{reference antenna}, and describe how the range and velocity of the subjects are estimated. In \secref{sec:az-el}, we extend the discussion to multiple receiver antennas, showing how the AZ and EL AoAs are computed.

\subsection{Range and Doppler information} \label{sec:rdmap}

Next, we show how to extract the range and velocity information from the received signal, focusing on the reference antenna. The signal reflected by a target is an attenuated version of the transmitted waveform with a delay $\tau$ that depends on the distance between the target and the radar and on their relative radial velocity. 

Denoting by $c$ the speed of light, and letting $R$ and $v$ respectively be the range and velocity of the target with respect to the radar device, the reflected signal delay is
\begin{equation}\label{eq:2d-delay}
	\tau = \frac{2(R + v t)}{c}.
\end{equation}
After mixing and sampling, the IF signal is expressed as~\cite{patole2017automotive}
\begin{equation}\label{eq:2d-signal}
	y(m, l) = \alpha \exp{\left[j \varphi(m, l)\right]} + \mathrm{w}(m, l),
\end{equation}
where $m$ and $l$ represent the sampling indices along the fast and slow time, respectively, $\alpha$ is a coefficient accounting for the attenuation effects due to the antenna gains, path loss and radar cross section (RCS) of the target and $\mathrm{w}(m,l)$ is a Gaussian noise term. The phase $\varphi(m,l)$ depends on the fast time and slow time sampling indices. By neglecting the terms giving a small contribution, an approximate expression for $\varphi(m,l)$ is written by introducing the quantities \mbox{$f_{d}=2f_o v/c$} and \mbox{$f_{b}=2\zeta R/c$}, which respectively represent the Doppler frequency and the \emph{beat} frequency of the reflected signal,
\begin{equation}\label{eq:2d-phase}
	\varphi(m,l) \approx 2\pi \left[\frac{2f_o R}{c} + f_{d} l T_{\mathrm{rep}} + \left(f_{d} + f_{b} \right)mT_f\right].
\end{equation}
Samples of $y(m, l)$ can be arranged into an \mbox{$M\times L$} matrix containing all the information provided by a single antenna for a given time frame. The frequency shifts of interest, which reveal the range and velocity of each reflector, can be extracted after applying a bi-dimensional discrete Fourier transform (DFT) along the fast time and slow time dimensions, followed by taking the square magnitude of each obtained complex value.
The result of this process is often referred to as radar \textit{range-Doppler map} (RD), and represents the received power distribution along the range of distances and velocities of interest.

The detection of the main reflecting points is performed using the \textit{cell-averaging constant false alarm rate} (CA-CFAR) algorithm on the \mbox{range-Doppler} maps \cite{richards2010principles}, which consists in applying a dynamic threshold on each RD value (or \textit{bin}), depending on the power of nearby \emph{training} values.  The use of an adaptive threshold introduces sparsity in the resulting set of detected points, as a point is retained (i.e., selected) only if its power is sufficiently larger than the average power of its neighbors.  

In addition, a processing step is required to remove the reflections from static objects, i.e., the \emph{clutter}. This operation is performed using a \textit{moving target indication} (MTI) high pass filter that removes the reflections with Doppler frequency values close to zero~\cite{richards2010principles}.

The detection and MTI processing steps return a sparse RD map containing $N^{\mathrm{det}}$ detected reflecting points: the position of each value along the fast time reveals the corresponding frequency in the IF signal \mbox{$f_{d} + f_{b} \approx f_{b}$}, while the peak along the slow time reveals the Doppler frequency $f_{d}$. 
For each detected point, the \textit{observed} desired quantities are then expressed as follows (we indicate with the symbol $\Delta$ the corresponding resolution)
\begin{equation}\label{eq:range-est}
	\tilde{R} = \frac{f_{b} c}{2 \zeta}, \quad \Delta\tilde{R} = \frac{c}{2B},
\end{equation}
\begin{equation}\label{eq:vel-est}
	\quad \tilde{v} = \frac{f_{d} c}{2f_o}, \quad \Delta\tilde{v} = \frac{c}{2f_oLT_{\mathrm{rep}}N_{\mathrm{TX}}}.
\end{equation}
Additionally, from the RD map we obtain the reflected, received power from each detection, denoted by $P^{\mathrm{RX}}$.

\subsection{Azimuth and Elevation angles estimation}\label{sec:az-el}

The complex-valued RD map of the radar illuminated range, before taking the square  magnitude, is computed at all the receiving antenna elements, and presents a different phase shift at each antenna, due to its different distance from the target. This fact is referred to as \textit{spatial diversity} of the receiver array, and can be exploited to estimate the azimuth and elevation angles of the targets.

Denote by $d$ the distance between two subsequent antennas along the azimuth and elevation dimensions and by $\psi_{\mathrm{AZ}}$ and $\psi_{\mathrm{EL}}$ the corresponding experienced phase shifts, respectively. Moreover, let $\theta$ and $\phi$ be the AZ and EL angles of a reflecting point, while $\lambda = c/f_o$ is the base wavelength of the transmitted chirps. The following relations hold
\begin{equation}\label{eq:ps-geom}
	\begin{split}
		\psi_{\mathrm{AZ}} & \approx \frac{2\pi}{\lambda} d \cos \phi \sin \theta,\\
		\psi_{\mathrm{EL}} & \approx  \frac{2\pi}{\lambda} d \sin \phi .
	\end{split}
\end{equation}

To compute the phase shift values, two DFTs across the samples taken at the azimuth and elevation antennas in the virtual receiver array are computed, extracting the peak positions similarly to what described in \secref{sec:rdmap} for beat and Doppler frequency.
Finally, the Cartesian coordinates of each detected point are obtained using \eq{eq:ps-geom} as
\begin{equation}\label{eq:conv-meas}
	\begin{split}
		\tilde{x} & = \tilde{R} \cos \phi \sin \theta = \tilde{R} \frac{\lambda\psi_{\mathrm{AZ}}}{2\pi d}, \\
		\tilde{y} & = \sqrt{\tilde{R}^2 - \tilde{x}^2 - \tilde{z}^2},\\
		\tilde{z} & = \tilde{R} \sin \phi  = \tilde{R} \frac{\lambda\psi_{\mathrm{EL}}}{2\pi d}.
	\end{split}
\end{equation}
The vector describing a single detected reflecting point, $\boldsymbol{p}_r$, \mbox{$r=1, \dots, N^{\mathrm{det}}$}, has five components, containing the information on its Cartesian coordinates, its velocity and the reflected power: \mbox{$\boldsymbol{p}_r = \left[\tilde{x}_r, \tilde{y}_r, \tilde{z}_r, \tilde{v}_r, P^{\mathrm{RX}}_r\right]^T $}.

\section{System Design}\label{sec:system}
\begin{figure*}[t!]
	\begin{center}   
		\includegraphics[width=13cm]{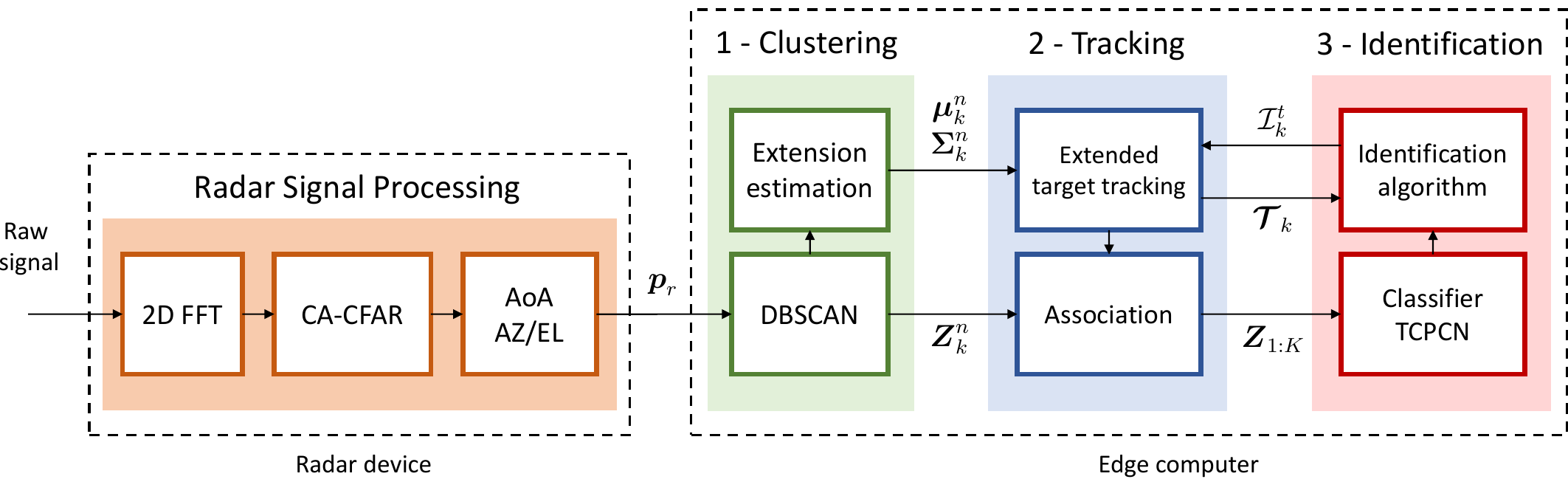} 
		\caption{Block diagram of the proposed signal processing workflow: the raw radar data is processed on the radar device, extracting the sparse point-cloud representation of the environment, i.e., points $\boldsymbol{p}_r$, then (1) a clustering module groups the points $\boldsymbol{p}_r$ into the contributions from the different targets and estimates their position and extension, (2-3) tracking, data association and identification are {\it jointly} performed through an identification algorithm.
		}
		\label{fig:workflow}
	\end{center}
\end{figure*}
The proposed system operates on discrete time steps, indicized by variable $k$, whose duration corresponds to the radar inter frame time $\Delta t$. At each frame, a set of $N_k^{\mathrm{det}}$ reflecting points $\boldsymbol{p}_r$ are obtained through the signal processing steps of \secref{sec:mmwave-model}. Our system sequentially performs the following operations on such points, see \fig{fig:workflow}.
\begin{enumerate}
	\item \textbf{Clustering and extension observation:} a density-based clustering algorithm is used to group the points detected by CA-CFAR into several clusters, each corresponding to a different subject present in the environment, see \secref{sec:clustering}. The points associated with the different targets are then used to obtain \textit{observations} of the subject's state, which according to our design includes his/her Cartesian position and \textit{extension} in the horizontal plane ($x-y$). The extension is modeled as an ellipse, that is determined by the spread (covariance) of the points in each cluster, \secref{sec:shape-det}.
	\item \textbf{Tracking and data association:} a CM-KF~\cite{lerro1993tracking} is used to estimate the position, velocity and extension of the subjects in a \mbox{multi-target} tracking (MTT) framework, processing the observations outputted by the previous step, \secref{sec:eot-model}. A set of trajectories, each corresponding to a human subject, are maintained and sequentially updated. The MTT association between new observations and trajectories is achieved using an approximation of the \textit{nearest-neighbors joint probabilistic data association} (NN-JPDA) algorithm, see \secref{sec:data-assoc}.
	\item \textbf{Identification:} a deep NN classifier is applied to a temporal sequence of $K$ subsequent \mbox{point-clouds} associated with each trajectory, with the objective of discerning among a set of $Q$ pre-defined subject identities. The employed NN is called \textit{temporal convolution point-cloud network} (TCPCN), and is inspired by the popular \mbox{PointNet} architecture used for 3D \mbox{point-cloud} classification and segmentation~\cite{qi2017pointnet}. TCPCN extends \mbox{PointNet} to the radar domain, by adding the velocity and received power information to the input and accounting for an additional block that handles the extraction of temporal features. Also, TCPCN is used in conjunction with an identification algorithm, which includes an exponential \mbox{moving-average} smoother and the Hungarian method, to jointly output a unique label for each trajectory: this combined use greatly improves the identification accuracy of the framework.
\end{enumerate}

\subsection{Point-cloud clustering -- DBSCAN}\label{sec:clustering}

\mbox{Density-based} clustering algorithms, as opposed to \emph{distance}-based ones, group input samples according to their local density. One of the most widely used algorithms belonging to this category is \mbox{DBSCAN}~\cite{ester1996density}, which has been successfully applied to cluster radar point clouds in~\cite{wagner2017radar, zhao2019mid, meng2020gait, pegoraro2020multiperson}. The algorithm operates a sequential scanning of all the data points, expanding a cluster until a certain density connectivity condition is no longer met. The algorithm takes two input parameters, $\varepsilon$ and $m_{\mathrm{pts}}$, respectively representing a radius around each point and the minimum number of other points that must be inside such radius to meet the density condition. DBSCAN is only applied to the $x-y$ components of the detected points $\boldsymbol{p}_r$, namely, the Cartesian coordinates on the horizontal plane, as the different body parts of a subject can have very different velocity and reflected power values. 
We denote by $\left\{\boldsymbol{Z}^n_k\right\}_{n=1, \dots, D_k}$ the $D_k$ clusters obtained at time step $k$ by grouping the $N_{k}^{\mathrm{det}}$ detected points. In principle, there should be a distinct cluster for each human subject present in the environment, but due to several phenomena such as noise, imperfect clutter cancellation and blockage of the signal, a subject can go undetected even for several consecutive frames. DBSCAN was chosen for the following reasons: it is an unsupervised algorithm, i.e., the number of clusters (subjects) does not have to be known beforehand, it has a noise rejection quality that, together with its \mbox{density-based} clustering mechanism, allows a reliable and automatic separation of the reflections from distinct subjects, it has a low computational complexity, of about $\mathcal{O} \left(N^{\mathrm{det}}_k\log N^{\mathrm{det}}_k\right)$.

\subsection{Subject Position and Extension Observations}\label{sec:shape-det}

Due to the high spatial resolution of mm-wave radars, human subjects are detected as clusters containing tens of reflecting points. In the literature, the typical approach to their tracking has been to ignore the spatial extension of the targets, considering them as ideal \mbox{point-shaped} reflectors. In the present work, given a cluster of points $\boldsymbol{Z}^n_k$ selected by the DBSCAN clustering algorithm at time $k$, we instead obtain an estimate of the extension of the subject in the \mbox{$x - y$} plane. As a first step, we define $\tilde{\boldsymbol{p}}_r = \left[\tilde{x}_r, \tilde{y}_r\right]^T$ and we normalize the received power values, $P_r^{\mathrm{RX}}$, of the detected points in $[0, 1]$. The spread of the points within each cluster around the cluster centroid provides a measure of the subject's extension. The centroid represents a noisy observation of the true position of the person, and is obtained as
\begin{equation}
	\boldsymbol{\mu}^n_k = \sum_{r : \tilde{\boldsymbol{p}}_r \in \boldsymbol{Z}^n_k} P_r^{\mathrm{RX}} \tilde{\boldsymbol{p}}_r,
\end{equation}
where $\boldsymbol{\mu}^n_k = [\mu^n_{x,k},\mu^n_{y,k}]^T$ and the received normalized powers $P_r^{\mathrm{RX}}$ act as weights. The covariance matrix, $\boldsymbol{\Sigma}^n_k$, contains information on the dimensions of the ellipse representing the extension of cluster $n$, and is obtained through the weighted sample covariance estimator,
\begin{equation}
	\boldsymbol{\Sigma}^n_k = \sum_{r : \tilde{\boldsymbol{p}}_r \in \boldsymbol{Z}^n_k} P_r^{\mathrm{RX}}\left(\tilde{\boldsymbol{p}}_r - \boldsymbol{\mu}^n_k\right)\left(\tilde{\boldsymbol{p}}_r - \boldsymbol{\mu}^n_k\right)^T.
\end{equation}

The norms of the eigenvectors of matrix $\boldsymbol{\Sigma}^n_k$, denoted by $\tilde{\ell}^n_k$ and $\tilde{w}^n_k$ provide the axes lengths of the ellipse, while the orientation, $\tilde{\xi}^n_k$, has the same direction of the eigenvector corresponding to the largest eigenvalue of $\boldsymbol{\Sigma}^n_k$.

\subsection{Extended Object Tracking -- Converted Measurements Kalman Filter}\label{sec:eot-model}

With the tracking step, we perform a sequential estimation of the \textit{state} of the subjects present in the environment from their observed positions and extensions.
To this end, we use a set of \mbox{CM-KFs} to establish a \mbox{so-called} \emph{track} for each subject. A new KF model is initialized for each detected cluster in the first frame received by the radar, while in successive frames, the tracks are maintained through the KF \mbox{predict-update} steps~\cite{kalman1960new}.
We denote by $\mathcal{T}^t_k$ the track with index $t$ at time $k$, by $\boldsymbol{\mathcal{T}}_k$ the set of currently maintained tracks, i.e., \mbox{$\boldsymbol{\mathcal{T}}_k = \left\{\mathcal{T}^t_k\right\}_{t = 1, \dots, T_{k}}$}, and by $T_{k}$ its cardinality. We define the state of $\mathcal{T}^t_k$ as \mbox{$\boldsymbol{\mathrm{x}}^t_{k} = \left[x^t_{k}, y^t_{k},\dot{x}^t_{k}, \dot{y}^t_{k}, \ell^t_{k}, w^t_{k}, \xi^t_{k}\right]^T$}, which contains the true (and unknown) user's position ($x^t_{k}$ and $y^t_{k}$), velocity ($\dot{x}^t_{k}$ and  $\dot{y}^t_{k}$), extension ($\ell^t_{k}$ and $w^t_{k}$) and orientation angle ($\xi^t_{k}$). Each track is then defined as a tuple, \mbox{$\mathcal{T}^t_k = \left(\hat{\boldsymbol{\mathrm{x}}}_k^t, \boldsymbol{P}_k^t, \boldsymbol{Z}^{t}_{k-K+1:k}, \mathcal{I}_k^t\right)$}, containing respectively the current state estimate, $\hat{\boldsymbol{\mathrm{x}}}_k^t$, the associated error covariance matrix as computed by the KF, $\boldsymbol{P}_k^t$, the collection of the last $K$ clusters associated with the track, $\boldsymbol{Z}^{t}_{k-K+1:k}$, to be fed to the NN classifier, and an integer $\mathcal{I}_k^t$ representing an estimate of the identity of the associated subject, at time $k$.
The observation vector for a detected target $n$ at time $k$ is \mbox{$\boldsymbol{\mathrm{z}}^n_{k} = \left[\mu_{x,k}^n, \mu_{y, k}^n, \tilde{\ell}^n_{k}, \tilde{w}^n_{k}, \tilde{\xi}^n_{k}\right]^T$}.

The matching between any given cluster $n$ and a corresponding track $t$ ($n \leftrightarrow t$) is carried out using a specific procedure that will be detailed shortly in \secref{sec:data-assoc}. For the sake of a concise notation, for the remainder of this section we drop the indices $n$ and $t$, as the procedure that we describe next is carried out independently for each track (subject) once the matching $n \leftrightarrow t$ is performed.

Given the sequence of all collected measurements for a track up to time $k$, $\boldsymbol{\mathrm{z}}_{1:k}$, the state estimation is carried out using the \mbox{CM-KF}. This approach assumes a posterior Gaussian distribution of the state given the sequence of measurements, i.e., $p(\boldsymbol{\mathrm{x}}_k | \boldsymbol{\mathrm{z}}_{1:k}) = \mathcal{N}(\hat{\boldsymbol{\mathrm{x}}}_k, \boldsymbol{P}_k)$. To update $\hat{\boldsymbol{\mathrm{x}}}_k$ and $\boldsymbol{P}_k$, a KF recursion \cite{kalman1960new} is applied using the measurements transformed in Cartesian coordinates from \secref{sec:shape-det}.

The model of motion that is used by the Kalman filtering block is defined by two matrices, $\boldsymbol{F}$ and $\boldsymbol{H}$. $\boldsymbol{F}$ is the transition matrix, connecting the system state at time $k$, $\boldsymbol{\mathrm{x}}_k$, to that at time $k-1$, $\boldsymbol{\mathrm{x}}_{k-1}$. $\boldsymbol{H}$ is the observation matrix, which relates the observation vector $\boldsymbol{\mathrm{z}}_k$ to the true state $\boldsymbol{\mathrm{x}}_k$. Referring to \mbox{$\boldsymbol{u}_k\sim \mathcal{N}\left(\boldsymbol{0}, \boldsymbol{Q}\right)$} and \mbox{$\boldsymbol{r}_k\sim \mathcal{N}\left(\boldsymbol{0}, \boldsymbol{R}_k\right)$} as the process noise and observation noise, respectively, a dynamic model of the system is 
\begin{equation}\label{eq:stateup}
	\boldsymbol{\mathrm{x}}_k = \boldsymbol{F}\boldsymbol{\mathrm{x}}_{k-1} + \boldsymbol{u}_k,
\end{equation}
\begin{equation}\label{eq:obs-state}
	\boldsymbol{\mathrm{z}}_k = \boldsymbol{H}\boldsymbol{\mathrm{x}}_k + \boldsymbol{r}_k.
\end{equation}
Denoting by $\mathrm{blkdiag}[\boldsymbol{A}, \boldsymbol{B}]$ the block diagonal matrix with blocks given by matrices $\boldsymbol{A}$ and $\boldsymbol{B}$, we have
\begin{equation}\label{eq:f-mat}
	\boldsymbol{F} = \mathrm{blkdiag}
	\left[
	\left[
	\begin{array}{cc}
		1 & \Delta t \\ 0 & 1
	\end{array}
	\right] \otimes\boldsymbol{I}_2 , \boldsymbol{I}_3
	\right],
\end{equation}
and
\begin{equation}\label{eq:h-mat}
	\boldsymbol{H} = \left[
	\begin{array}{ccc}
		\boldsymbol{I}_2 & \boldsymbol{0}_{2\times 2}  & \boldsymbol{0}_{2\times 3}  \\ 
		\boldsymbol{0}_{3\times 2} & \boldsymbol{0}_{3\times 2} & \boldsymbol{I}_3 
	\end{array}
	\right],
\end{equation}
where $\boldsymbol{I}_n$ is an $n\times n$ identity matrix, $\boldsymbol{0}_{n\times m}$ is an $n\times m$ all-zero matrix and $\otimes$ refers to the Kronecker product between matrices.

We assume the process noise $\boldsymbol{u}_k $ is due to a random acceleration $a_k$ that follows a Gaussian distribution with $0$ mean and variance $\sigma_a^2$, i.e., $a_k \sim \mathcal{N}(0, \sigma_a^2)$, leading to \mbox{$\boldsymbol{u}_k = \boldsymbol{g}a_k$} with $\boldsymbol{g} = \left[\Delta t^2/2, \Delta t\right]^T$.
The process noise covariance matrix is obtained as
\begin{equation}\label{eq:q-mat}
	\boldsymbol{Q} = \mathrm{blkdiag} \left[
	\sigma_a^2\boldsymbol{g}\boldsymbol{g}^T \otimes \boldsymbol{I}_2, \mathrm{diag}\left( \sigma^2_{\ell}, \sigma^2_{w}, \sigma^2_{\xi} \right)
	\right],
\end{equation}
with $\sigma^2_{\ell}, \sigma^2_{w}, \sigma^2_{\xi}$ being the constant process noise variances on the extension- and \mbox{orientation-related} coordinates of the state.
The observation noise has covariance matrix given by
\begin{equation}\label{eq:r-mat}
	\boldsymbol{R}_k = \mathrm{blkdiag}\left[
	\boldsymbol{R}'(\boldsymbol{\mathrm{x}}_{k}), \mathrm{diag}\left( \sigma^2_{\tilde{\ell}}, \sigma^2_{\tilde{w}}, \sigma^2_{\tilde{\xi}}  \right)
	\right],
\end{equation}
with $\sigma^2_{\tilde{\ell}}, \sigma^2_{\tilde{w}}, \sigma^2_{\tilde{\xi}} $ being the constant observation noise variances on the extension- and \mbox{orientation-related} coordinates of the state. For what concerns $\boldsymbol{R}^\prime$, as radar measurements are obtained in polar coordinates, and then converted to the Cartesian space using \eq{eq:conv-meas}, the measurement covariance matrix is \mbox{time-varying} as it depends on the current target's position. The \mbox{sub-matrix} $\boldsymbol{R}^\prime$ accounts for the uncertainty in the Cartesian position observations, reflecting that an error on the AoA causes a higher uncertainty in Cartesian coordinates as the distance of the subject increases, due to the \mbox{non-linear} mapping between polar and Cartesian coordinates. In setting the uncertainty parameters for the measurements, we use a constant measurement covariance in polar coordinates, \mbox{$\boldsymbol{R}_{\mathrm{pol}} = \mbox{diag}\left(\sigma^2_{R}, \sigma^2_{\theta}\right)$}, where $R$ and $\theta$ are the distance and azimuth AoA, respectively introduced in \secref{sec:rdmap} and \secref{sec:az-el}. Hence, we use the transform \mbox{$\boldsymbol{R}^\prime(\boldsymbol{\mathrm{x}}_{k}) = \boldsymbol{J}_{|\boldsymbol{\mathrm{x}}_{k}}\boldsymbol{R}_{\mathrm{pol}}\boldsymbol{J}^T_{|\boldsymbol{\mathrm{x}}_{k}}$}, where $\boldsymbol{J}_{|\boldsymbol{\mathrm{x}}_{k}}$ is the Jacobian matrix of the conversion between polar and Cartesian coordinates, computed using the polar representation of the true subject state, $\boldsymbol{\mathrm{x}}_{k}$, which we approximate with $\boldsymbol{\mathrm{x}}_{k} \approx \boldsymbol{H}\hat{\boldsymbol{\mathrm{x}}}_{k-1}$. Although it can be seen that our conversion to Cartesian coordinates is biased, we remark that employing the unbiased conversion proposed in~\cite{bordonaro2014decorrelated} did not lead to significant improvements. Note that, by the structure of the model matrices in \eq{eq:f-mat} and \eq{eq:h-mat}, the kinematic part of the subject state and the extension part are entirely decoupled and do not interact during the \mbox{CM-KF} operations.

As a final remark about the KF model, with our approach the extension of the subject is explicitly accounted for as part of the state, fitting the \mbox{point-clouds} with ellipses, similarly to~\cite{gennarelli2015multiple}. Although other approaches exist, such as using random matrices~\cite{koch2008bayesian, feldmann2010tracking}, we found that our method leads to more accurate and meaningful extension estimates of the target's shape, due to the fast variability of radar \mbox{point-clouds}. 

\subsection{Data Association -- NN-CJPDA}\label{sec:data-assoc}

\begin{figure*}[t!]
	\begin{center}   
		\includegraphics[width=13cm]{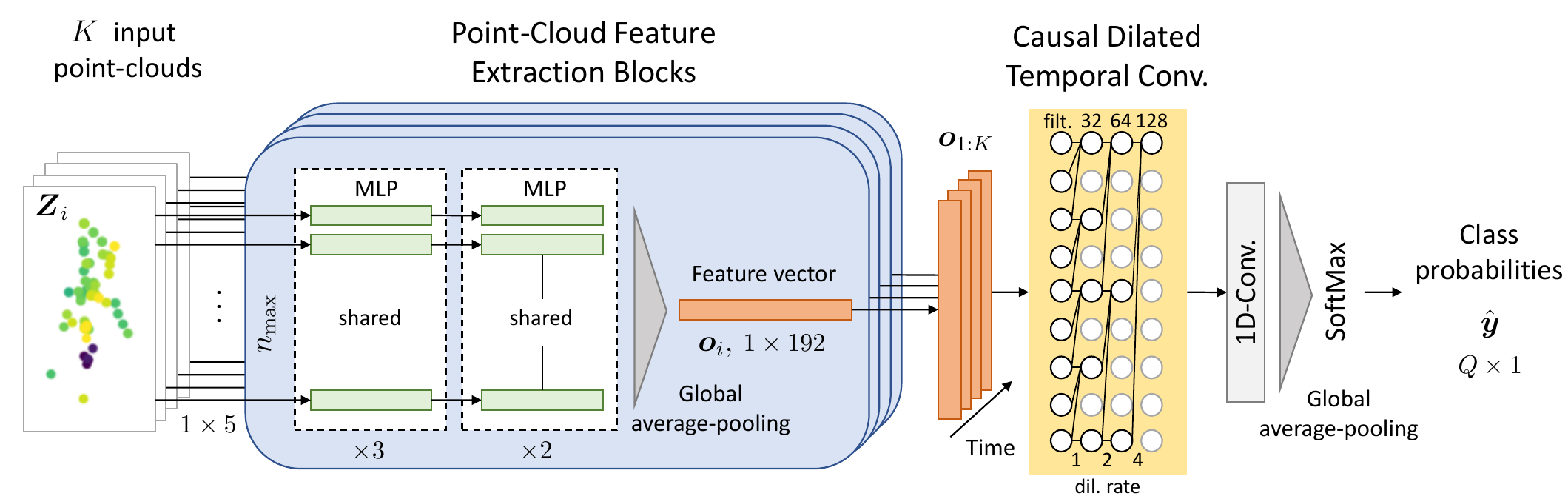} 
		\caption{TCPCN -- proposed \mbox{DL-based} classifier for subject identification: \textit{(i)} a \textit{point-cloud block} is applied to each individual time step to extract a feature vector, \textit{(ii)} causal dilated convolutions are used to learn the temporal patterns in the sequence of feature vectors.}
		\label{fig:classifier}
	\end{center}
\end{figure*}

The association between new observations and tracks is needed \textit{(i)} to correctly update the tracks with the observations generated by the corresponding subjects in a \mbox{multi-target} scenario, \textit{(ii)} to correctly collect the sequence of the past $K$ \mbox{point-clouds} associated with each subject, $\boldsymbol{Z}^{t}_{k-K+1:k}$.  

To match tracks $t$ to clusters $n$ ($n \leftrightarrow t$), we use the \emph{nearest-neighbors joint probabilistic data association} (\mbox{NN-JPDA}) scheme. This method consists in computing the probability of each possible association between the $D_k$ new clusters and the previous $T_k$ tracks. These probabilities are then arranged into a $D_k \times T_{k-1}$ matrix of scores, $\boldsymbol{\Gamma}$, and the final assignment is done considering the association leading to the maximum overall probability, computed using the Hungarian algorithm~\cite{kuhn1955hungarian}. The Hungarian algorithm uses the score matrix as input and solves the problem of pairing each track with only one cluster while maximizing the total score, entailing an overall complexity \mbox{$\mathcal{O}((T_{k-1} D_k)^3)$}.

To compute the probability of each match, i.e., the elements of matrix $\boldsymbol{\Gamma}$, we consider the widely adopted JPDA logic, using the approximate version of~\cite{fitzgerald1986development} called \mbox{cheap-JPDA} (CJPDA). Exploiting the fact that the kinematic, extension and orientation parts of the state are decoupled in our framework, we apply CJPDA only using the kinematic state, as extension and orientation are more unreliable and could lead to association errors. Hence, in the following we refer to the kinematic part of the KF vectors and matrices only, i.e., to the components related to the Cartesian position and velocity of the targets. 

The score matrix $\boldsymbol{\Gamma}$ is computed as follows (the time index $k$ is omitted for a simpler notation). First, for all track-detection pairs the quantity $G_{nt}$ is computed, which is proportional to the Gaussian function expressing the likelihood that observation $n$ is produced by the subject corresponding to track $t$ 
\begin{equation}
	G_{nt} = \frac{1}{\sqrt{\det \boldsymbol{S}_{nt}}} \exp{\left[ -\frac{1}{2}\boldsymbol{\nu}_{nt}^{T}\boldsymbol\left(\boldsymbol{S}_{nt}\right)^{-1} \boldsymbol{\nu}_{nt}\right]},
\end{equation}
where \mbox{$\boldsymbol{\nu}_{nt} = \hat{\boldsymbol{\mathrm{x}}}^t - \boldsymbol{H}\boldsymbol{\mathrm{z}}^n$} is the \emph{innovation} brought by measurement $\boldsymbol{\mathrm{z}}^n$ to the kinematic state of track $t$, $\hat{\boldsymbol{\mathrm{x}}}^t$, and \mbox{$\boldsymbol{S}_{nt} = \boldsymbol{H}\boldsymbol{P}^t\boldsymbol{H}^{T} + \boldsymbol{R}$} is its covariance matrix, obtained as part of the KF recursion. Second, the association probabilities for each \mbox{track-detection} pair are computed following~\cite{fitzgerald1986development}, as
\begin{equation}
	\Gamma_{nt} = \frac{G_{nt}}{\sum_{t=1}^{T_{k-1}} G_{nt} + \sum_{n=1}^{D_k} G_{nt} - G_{nt} + \beta},
\end{equation}
where the bias term $\beta$ accounts for the possibility that no measurement is a good match for a specific track and is connected with the probability of missed detection. In this work, $\beta$ is empirically set to $\beta=0.01$, preventing the association of track-detection pairs with a low $G_{nt}$ score. 

\subsection{Track management}
\label{sec:trackman}

The proposed system is robust to subjects that randomly appear on and disappear from the monitored space: these events may happen due to blockage of the radar signal at any point in time, or because the subject has moved in or out of the radio range. Blockage is a frequent problem in \mbox{mm-wave} propagation and it happens frequently in \mbox{multi-target} scenarios, as users may block the radio signal with their own body. To deal with undetected subjects and new cluster detections which cannot be reliably associated with any existing track, while keeping the complexity of the system as low as possible, we follow a \mbox{so-called} $\texttt{m}/\texttt{n}$ logic. In detail, a track is maintained if it received a match with any of the clusters detected by DBSCAN for at least $\texttt{m}$ out of the last $\texttt{n}$ frames. Similarly, cluster detections that are not associated with any existing track are initialized as new trajectories if they are detected for at least $\texttt{m}$ out of the last $\texttt{n}$ frames. In addition, to avoid tracks to merge when the subjects move too close to one another, the \mbox{inter-track} proximity is monitored.
If the estimated Euclidean distance\footnote{Obtained as $d(\mathcal{T}^t_k,\mathcal{T}^{t^\prime}_k) = ((x_k^t-x_k^{t^\prime})^2+(y_k^t-y_k^{t^\prime})^2)^{1/2}$.} between any two tracks $\mathcal{T}^t_k$ and $\mathcal{T}^{t^\prime}_k$ becomes smaller than the DBSCAN radius parameter, $\varepsilon$, we remove the track having the largest determinant of the estimated error covariance, i.e., $\arg\!\max_{j \in \left\{t, t^\prime\right\}} (\det \boldsymbol{P}^j_k)$.

\subsection{Point-cloud pre-processing}\label{sec:pc-preproc}

The point cloud sequence $\boldsymbol{Z}_{k-K+1:k}^{t}$ obtained from each \mbox{CM-KF} track is pre-processed before being sent to the NN classifier. The features of the points are standardized by subtracting their mean value and dividing by their empirical standard deviation. Moreover, the \mbox{point-clouds} must contain a fixed number of points before being sent to the TCPCN, as the latter is a feed forward neural network processing fixed size input vectors. We chose to limit the maximum number of points for a single time step to \mbox{$n_{\max}=100$}. In case the number of points is greater than such maximum value, we randomly sample $n_{\max}$ points from the \mbox{point-cloud} without repetitions, in case there are fewer points than $n_{\max}$, some of the points are randomly repeated to reach the maximum value. The choice of $n_{\max}$ was made by analyzing the distribution of the number of detected points for different human subjects and empirically picking a suitable value: the selected $n_{\max}$ suffices to contain the \mbox{point-clouds} of all users in almost every frame in our experiments. Also, due to blockage and clutter, a subject may go undetected, especially in a \mbox{multi-target} scenario. If this occurs, the \mbox{point-cloud} data for the current frame is not collected for the blocked user and, in turn, is not sent to the NN classifier. A missed detection persisting over multiple radio frames may make the sequence of temporal features extracted for a subject by the NN less representative of his/her movement, and may ultimately degrade the identification performance of the algorithm. To ameliorate this, we propose an identification algorithm that jointly considers the outputs of the tracking block and of the classifier, as detailed in \secref{sec:id-alg}.

Considering that the TCPCN classifier is applied consistently to every track $t$ at every time step $k$, in the following we simplify the notation denoting the \mbox{pre-processed} input \mbox{point-cloud} sequence $\boldsymbol{Z}_{k-K+1:k}^{t}$, of length $K$, by $\boldsymbol{Z}_{1:K}$.

\subsection{Identification -- Temporal Convolution Point-Cloud Network} \label{sec:classifier}

The proposed classifier is designed to extract meaningful features from a temporal sequence of \mbox{point-clouds}, which is obtained as a result of the detection and tracking steps. The proposed architecture includes two processing blocks, termed \textit{point-cloud} block (PC) and \textit{temporal convolution} block (TC), and we refer to the full neural network as \textit{temporal convolution point-cloud network} (TCPCN), see \fig{fig:classifier}.

\subsubsection{Point-cloud Block}
A number $K$ of identical (same weights) feature extraction blocks is applied to the standardized input point-clouds, $\boldsymbol{Z}_{i}$, $i=1,\dots,K$, of size $n_{\max} \times 5$, i.e., each composed of $n_{\max}$ reflecting points $\boldsymbol{p}_r$ (see \secref{sec:az-el}). Each of such blocks implements a function $f_{\boldsymbol{W}}(\cdot)$, obtained as the cascade of a \mbox{multi-layer} perceptron (MLP) \cite{goodfellow2016deep} followed by a global average pooling operation, where $\boldsymbol{W}$ is a set of weights to be learned. Each reflecting point $\boldsymbol{p}_r$ in point-cloud $\boldsymbol{Z}_{i}$ (a vector of size $1 \times 5$), is fed to the first MLP layer and is independently processed from all the other $n_{\max}$ points in $\boldsymbol{Z}_{i}$, by one of   $n_{\max}$ parallel branches. The MLPs located at the same depth share the same weights across all the points: there are $3$ \textit{fully-connected} (FC) layers with $96$ units followed by $2$ FC layers with $192$ units. Each FC layer applies a linear transformation of the input followed by an \mbox{exponential-linear} unit (ELU) activation function~\cite{clevert2015fast}. Batch normalization is used after each linear transformation~\cite{ioffe2015batch} and right before the following \mbox{non-linearity} (ELU). The output feature vector from the last MLP layer from each branch has size $1 \times 192$. Global average pooling reduces this set of features to a single feature vector, $\boldsymbol{o}_i = f_{\boldsymbol{W}}\left(\boldsymbol{Z}_{i}\right)$, of size $1 \times 192$, by taking the average of each element across all the $100$ parallel branches. The structure of function $f_{\boldsymbol{W}}(\cdot)$ is loosely inspired by the popular PointNet~\cite{qi2017pointnet}. The key aspect of $f_{\boldsymbol{W}}(\cdot)$ is that it uses functions that are invariant to the ordering of the input points, by sharing the weights of the MLP and using suitable pooling operations. This ensures robustness and generality, because point-clouds that only differ in how the points are ordered will result in the same output. We underline that our TCPCN significantly differs from PointNet as the latter is designed to perform end-to-end classification and segmentation of {\it dense} 3D point clouds, whereas our $f_{\boldsymbol{W}}(\cdot)$ performs feature extraction from {\it sparse} 5D point-clouds.

\subsubsection{Temporal Convolution Block}
The sequence of feature vectors $\boldsymbol{o}_{1:K} = \left\{f_{\boldsymbol{W}}\left(\boldsymbol{Z}_{i}\right)\right\}_{i=1:K}$, each of dimension $192$, is then fed to the TC block, which operates along the temporal dimension applying a function $h_{\boldsymbol{U}}(\cdot)$, where $\boldsymbol{U}$ is another set of weights.
To extract temporal features efficiently, $h_{\boldsymbol{U}}(\cdot)$ contains temporal convolutions, which are a type of convolutional neural network (CNN) layer \cite{goodfellow2016deep} where the input is convolved with a \mbox{uni-dimensional} filter (or \emph{kernel}) of learned weights in order to recognize temporal patterns. The output of the filters is organized into so called \emph{feature maps}, which become more and more complex and abstract with the depth of the layer.
In TCPCN we use \emph{causal dilated convolutions}~\cite{yu2015multi,oord2016wavenet}. This technique consists \textit{(i)}~in~masking the filters in such a way that neurons corresponding to a certain time step only depend on neurons corresponding to past time steps, i.e., they can not use future information, as done in~\cite{oord2016wavenet}, and \textit{(ii)}~in~applying the convolution filters skipping blocks of $\delta-1$ samples in the input, where $\delta$ is the \mbox{so-called} \emph{dilation rate}. Formally, denoting a feature map as $\boldsymbol{\mathrm{m}}$ and the filter as $\boldsymbol{\mathrm{k}}$, the output of a dilated convolution, $*_{\delta}$, between $\boldsymbol{\mathrm{m}}$ and $\boldsymbol{\mathrm{k}}$ is~\cite{yu2015multi},  
\begin{equation}
	\left(\boldsymbol{\mathrm{m}} *_{\delta} \boldsymbol{\mathrm{k}}\right)(s) = \sum_{i + \delta j = s} \boldsymbol{\mathrm{m}}(i)\boldsymbol{\mathrm{k}}(j).
\end{equation}
The standard discrete convolution is obtained for $\delta=1$. In the proposed TCPCN we employ $3$ temporal convolution layers with filters of dimension $3$ (also called \textit{kernel} dimension) and dilation rates of $1, 2$ and $4$, respectively. The applied filters are repeated along the feature vector components of the input, obtaining $32, 64$ and $128$ feature maps at each layer, respectively.

The last layer of TCPCN is a temporal convolution layer that maps the extracted temporal features onto $Q$ feature maps, each corresponding to one of the output classes. It applies a standard convolution with a kernel size of $3$ and it is followed by a global average pooling to group the information from each feature map and obtain a single vector of dimension $Q$. Finally, a $\mathrm{SoftMax}$ function is applied, defined for a generic vector $\boldsymbol{x}$ as \mbox{$\mathrm{SoftMax}(\boldsymbol{x})_i = e^{x_i}/\sum_j e^{x_j}$}. The vector outputted by this last layer is denoted by \mbox{$\hat{\boldsymbol{y}} = \mathrm{SoftMax}(h_{\boldsymbol{U}}\left(\boldsymbol{o}_{1:K}\right)) = \mathrm{TCPCN}\left(\boldsymbol{Z}_{1:K}\right)$} and its $q$-th elements represents the probability that the input \mbox{point-cloud} sequence belongs to class $q$.

\subsection{Classifier Training and Inference}
\label{sec:train-inf}

\subsubsection{Loss Function}
The loss function used is the categorical cross-entropy, which is a standard choice in classification problems \cite{goodfellow2016deep}. The CE compares the output of the last layer $\hat{\boldsymbol{y}}$ with the ground-truth identities of the subjects expressed in one-of-$Q$ representation, $\boldsymbol{y}$: $\mathcal{L}\left(\hat{\boldsymbol{y}}, \boldsymbol{y}\right) = - \sum_{q=1}^{Q}y_q \log(\hat{y}_q)$.

\subsubsection{Training}\label{sec:training}
To train TCPCN, we used the Adam optimizer with learning rate $\eta=10^{-4}$ \cite{goodfellow2016deep}. 
The process is stopped once the loss function computed on a validation set of data stops decreasing, a technique called \textit{early stopping}. Overfitting is a severe problem in the context of radar point-clouds: the high randomness of the detected points and the sensitivity to different environments make the learning task challenging, especially when generalization to unseen environments is required. To reduce overfitting, several strategies were utilized: \textit{dropout}~\cite{srivastava2014dropout} was applied to the output of the PC blocks, randomly dropping components of the feature vectors with probability $p_{\mathrm{drop}}=0.5$, an $L_2$ regularization cost~\cite{goodfellow2016deep} on all network weights was considered, with parameter \mbox{$\lambda_{L_2}=10^{-4}$}.
The selection of the hyperparameters was carried out using a \textit{greedy search} procedure.

\subsubsection{Inference}
During the inference (or prediction) step, TCPCN is used to obtain classification probabilities for each maintained track, $\mathcal{T}^t_k \in \boldsymbol{\mathcal{T}}_k$, in the current time step $k$. We denote by $\tilde{\boldsymbol{y}}_k^t \in [0, 1]^Q$ the vector that collects these probabilities. The prediction is carried out on a batch of $T_k$ point-cloud sequences in parallel with a single pass of the data through the network, jointly obtaining $\left\{\tilde{\boldsymbol{y}}_k^t\right\}_{\mathcal{T}^t_k \in \boldsymbol{\mathcal{T}}_k}$. Moreover, we apply weight quantization, \cite{jacob2018quantization}, to $8$ bit integer values to reduce even further the inference time and the memory cost of the model. It is worth noting that, due to the use of convolutions, TCPCN has a low number of parameters: in the PC block the weights are shared among the $n_{\max}$ parallel branches, while the TC block is a \emph{fully convolutional} neural network, with no fully connected layers.
Fully convolutional networks are typically very fast in terms of training and inference time compared to fully connected or recurrent neural networks and have fewer parameters (further analysis is carried out in \secref{sec:sota-comp}).

\subsection{Identification algorithm} \label{sec:id-alg}

After obtaining the output probabilities for each track from TCPCN, several problems still have to be tackled: \textit{(i)} obtaining stable classifications, robust to the fact that subjects may turn or move in unpredicted ways which do not carry their typical movement signature, \textit{(ii)} finding a method to compensate for the missing frames when subjects go undetected, which can cause classification errors, \textit{(iii)} dealing with the uniqueness of the subject identities, as classifying the subjects independently and solely based on $\tilde{\boldsymbol{y}}_k^t$ may lead to assigning the same identity to multiple targets. To address these problems, we devised the procedure detailed in \alg{alg:labeling}, which uses both the output of the tracking procedure and the classification probabilities provided by TCPCN to estimate the identities of the subjects in a stable and reliable way.
The procedure acts as follows.
\begin{enumerate}
	\item At the first time step $k=1$, a vector $\boldsymbol{y}^t_1$ of size $Q$ is initialized for each track $\mathcal{T}^t_1 \in \boldsymbol{\mathcal{T}}_1$, with all components equal to $1/Q$. $\boldsymbol{y}^t_1$ represents a stabilized vector of probabilities for each track.
	\item At the generic time step $k>1$, $\boldsymbol{y}^t_k$ is updated using \alg{alg:labeling}, according to one of the two following rules: 
	\begin{enumerate}
		\item if track $\mathcal{T}^t_k$ was detected in the most recent $K/2$ time-steps (line 1), TCPCN is applied to the corresponding sequence of \mbox{point-clouds}, obtaining the probability vector $\tilde{\boldsymbol{y}}^t_k$ (line 2). Hence, an exponentially weighted moving average procedure (line 6) is applied to mediate between the previous stable estimate $\boldsymbol{y}^t_{k-1}$ and the newly computed one $\tilde{\boldsymbol{y}}^t_k$, obtaining a new stable estimate $\boldsymbol{y}^t_{k}$ (normalized so that its elements sum to one, see line 7).
		\item if track $\mathcal{T}^t_k$ was not detected in at least one of the most recent $K/2$ time steps (line 8), $\boldsymbol{y}^t_{k}$ is obtained as $\gamma \boldsymbol{y}^t_{k-1}$ with $\gamma < 1$ (line 9). In this way, we maintain the last reliable identification, but we progressively lower the confidence that we put on it over time. Note that after this step $\boldsymbol{y}^t_{k}$ does not longer resemble a probability distribution, as the sum of its elements is smaller than one.
	\end{enumerate}
	\item To assign identities to subjects without repetitions, we build a matrix of scores $\boldsymbol{Y}_k$ with all vectors $\boldsymbol{y}^t_{k}$ belonging to each track (line 11). We compute the best assignment of the identities using the Hungarian algorithm on $\boldsymbol{Y}_k$, which guarantees that the joint maximum score is attained with a \mbox{one-to-one} mapping (line 13).
\end{enumerate}
To avoid associating a label to a track if the corresponding probability is very low, in the identification process we use a slightly modified version of the Hungarian algorithm, which behaves as follows: first, we compute the associations using the standard Hungarian algorithm. Hence, if the probability of a certain association is below $p_{\mathrm{conf}}=0.1$, we set the identity of the considered track to {\it unknown}. In \alg{alg:labeling}, this modified Hungarian algorithm is indicated as Hungarian$({\boldsymbol{Y}}_k, p_{\mathrm{conf}})$ to highlight that the result is a function of the score matrix $\boldsymbol{Y}_k$ and of the confidence threshold $p_{\mathrm{conf}}$. 

With \alg{alg:labeling}, we jointly exploit the information from the classifier (vector $\tilde{\boldsymbol{y}}^t_{k}$) and the tracking step ($\boldsymbol{\mathcal{T}}_k^{(s)}$) to improve the identification performance. 

\alg{alg:correction} deals with errors in the tracking procedure, using the identity information available for each track. Tracking or association errors may happen during a blockage event involving two subjects (\textit{blocker} and \textit{blocked} in the following): for example a blocked subject may be erroneously associated when he/she becomes detectable again while being close to the blocker. These errors are dynamically corrected by analyzing the output of \alg{alg:labeling}. Specifically, when the identity of a track $\mathcal{T}^t_k$ changes, we assume that this is an indication of a tracking error of the \mbox{above-mentioned} type (see line 2 of \alg{alg:correction}). In this case, this track is removed from the set of tracks that are maintained (line 4). At the same time, a new track $\mathcal{T}^j_k$ is initialized using the new identity $\mathcal{I}_{k}^t$, a new track index $j$ (not yet used) and the current variables (state and covariance) associated with the old track $\mathcal{T}^t_k$ at time $k$ (line 3). The new track $\mathcal{T}^j_k$ is then added to the set of maintained tracks (line 4). Note that, the memory $\boldsymbol{Z}_{k-K+1:k}^{t}$ (past frames) is not attached to the new track, which is started anew. 



\begin{algorithm}[t!]
	\caption{Joint identification at time step $k$.}
	\label{alg:labeling}
	\begin{algorithmic}[1]
		\REQUIRE Current set of tracks, $\boldsymbol{\mathcal{T}}_k$, smoothing parameter, $\rho$, decay parameter, $\gamma$.
		\ENSURE Identities $\mathcal{I}_{k}^t, \quad  \forall \mathcal{T}^t_k \in \boldsymbol{\mathcal{T}}_k$.
		\STATE Set $\boldsymbol{\mathcal{T}}_k^{(s)} = \left\{\mathcal{T}^t_k \in \boldsymbol{\mathcal{T}}_k \mbox{ s.t. } \mathcal{T}^t_k \mbox{ det. in the last } K/2 \mbox{ frames}\right\}$
		\STATE $\left\{\tilde{\boldsymbol{y}}^t_k\right\}_{\mathcal{T}^t_k \in \boldsymbol{\mathcal{T}}_k^{(s)}} \leftarrow \mathrm{TCPCN}\left(\left\{\boldsymbol{Z}_{k-K+1:k}^{t}\right\}_{\mathcal{T}^t_k \in \boldsymbol{\mathcal{T}}_k^{(s)}}\right)$
		\STATE Initialize $\boldsymbol{Y}_k = \boldsymbol{0}_{ T_k \times  Q}$
		\FOR{$\mathcal{T}^t_k \in \boldsymbol{\mathcal{T}}_k$}
		\IF{$\mathcal{T}^t_k \in \boldsymbol{\mathcal{T}}_k^{(s)}$}
		\STATE $\boldsymbol{y}_k^t \leftarrow \left(1 - \rho\right)\tilde{\boldsymbol{y}}_k^t + \rho\boldsymbol{y}_{k-1}^t$ 
		\STATE normalize $\boldsymbol{y}_k^t$ 
		\ELSE
		\STATE $\boldsymbol{y}_k^t \leftarrow \gamma \boldsymbol{y}_{k-1}^t$ 
		\ENDIF
		\STATE $\left(\boldsymbol{Y}_k\right)_{t, :} \leftarrow\boldsymbol{y}_k^t$
		\ENDFOR
		\STATE $\mathcal{I}_{k}^t$ $\leftarrow$ Hungarian$(\boldsymbol{Y}_k, p_{\mathrm{conf}}), \, \forall \, \mathcal{T}^t_k \in \boldsymbol{\mathcal{T}}_k$ 
	\end{algorithmic}
\end{algorithm}
\begin{algorithm}[t!]
	\caption{Tracking error correction at time step $k$.}
	\label{alg:correction}
	\begin{algorithmic}[1]
		\REQUIRE Current set of tracks $ \boldsymbol{\mathcal{T}}_k$.
		\ENSURE Updated set of tracks $ \boldsymbol{\mathcal{T}}^\prime_k$.
		\FOR{$\mathcal{T}^t_k \in \boldsymbol{\mathcal{T}}_k$}
		\IF{$\mathcal{I}_{k}^t \neq \mathcal{I}_{k-1}^t $}
		\STATE initialize new track $\mathcal{T}^{j}_k$ using $\boldsymbol{\mathrm{x}}^t_{k}, \boldsymbol{P}^t_{k}$ and $\mathcal{I}_{k}^t$ 
		\STATE $\boldsymbol{\mathcal{T}}^\prime_k \leftarrow \left\{\boldsymbol{\mathcal{T}}_k \setminus \mathcal{T}^t_k\right\} \cup \{\mathcal{T}^{j}_k\}$
		\ENDIF
		\ENDFOR
	\end{algorithmic}
\end{algorithm}

\section{Experimental results}\label{sec:results}

\begin{figure*}[t!]
	\begin{center}   
		\centering
		\subcaptionbox{Jetson board (left) and radar  (right).\label{fig:jet-rad}}[5.5cm]{\includegraphics[width=4.6cm]{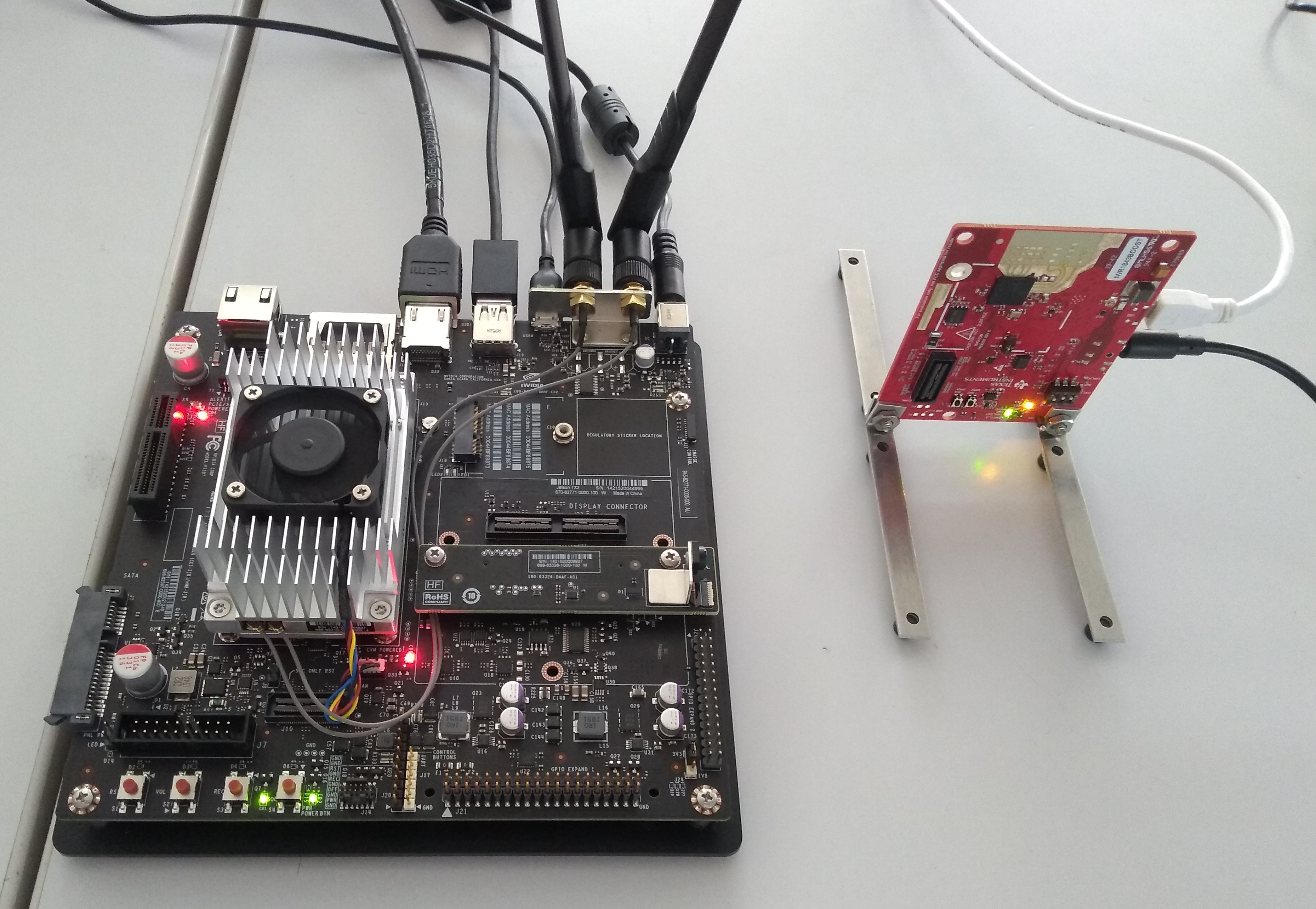}}
		\subcaptionbox{The mounted setup in the test room.\label{fig:jet-ra-_mount}}[5.5cm]{\includegraphics[width=3.05 cm]{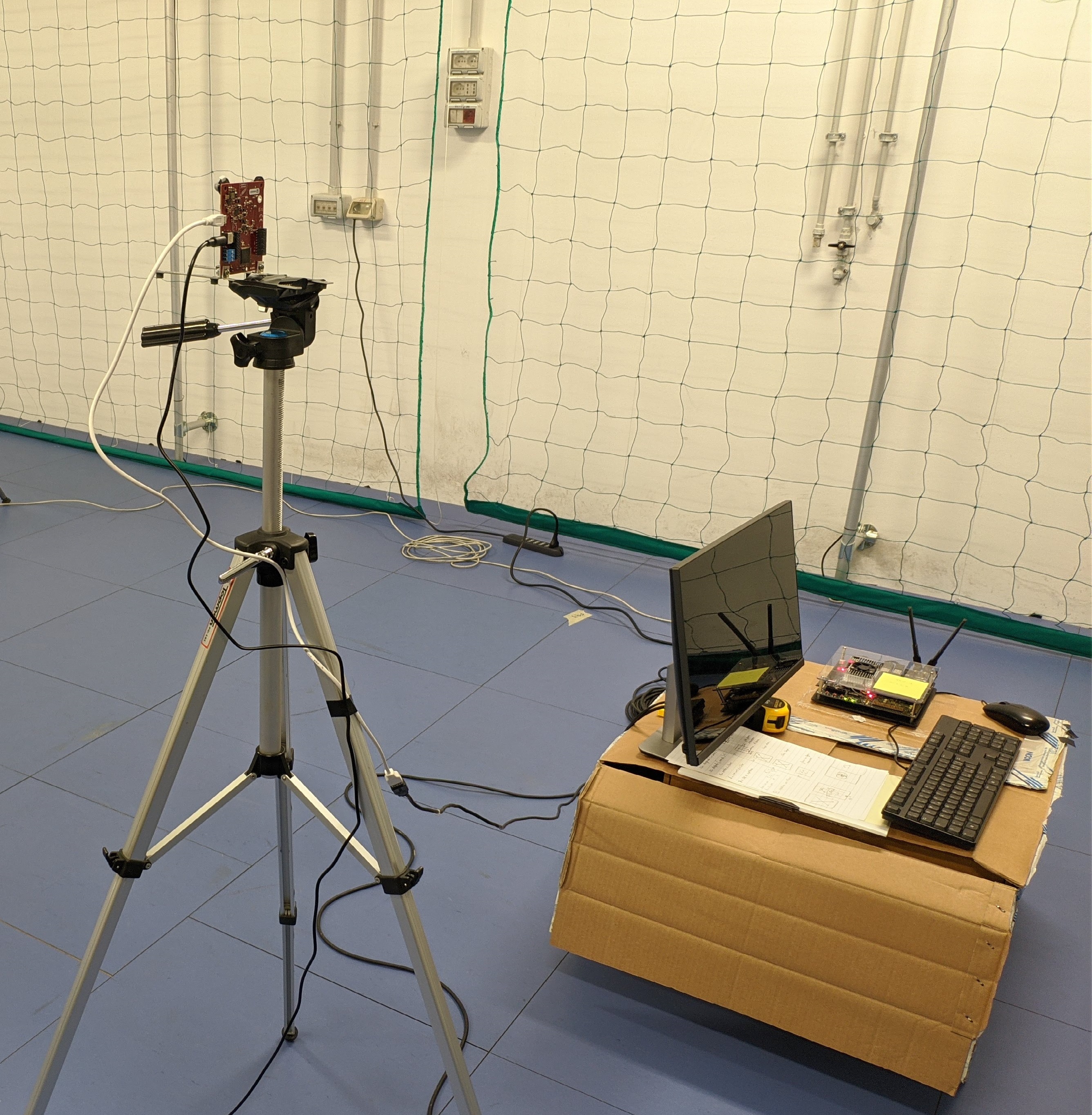}}
		\subcaptionbox{Three subjects walking in the test room.\label{fig:subjects}}[5.8cm]{\includegraphics[width=5.5cm]{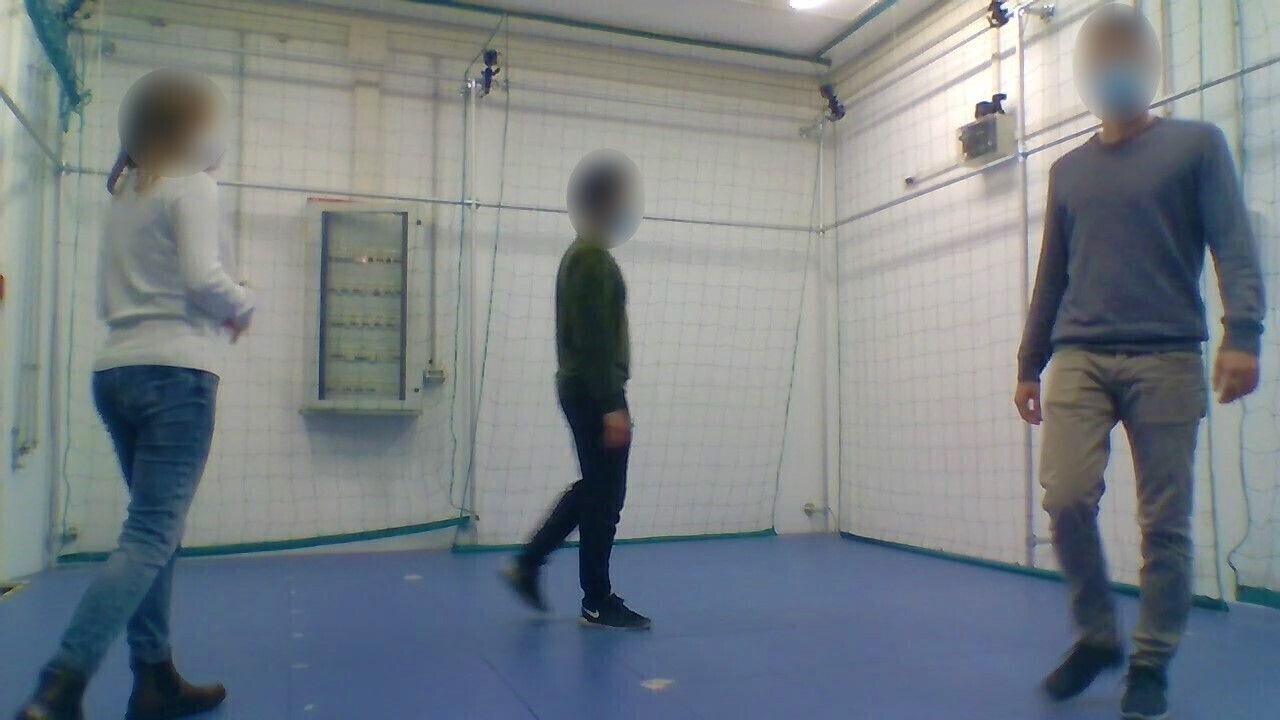}}
		\caption{Overview of the experimental setup.}
		\label{fig:setup}
	\end{center}
\end{figure*}

In this section, we present results obtained by evaluating our tracking and identification method on
\begin{enumerate}
	\item the mmGait dataset described in \mbox{\cite{meng2020gait}}, available at \url{https://github.com/mmGait/people-gait}  (\mbox{\secref{sec:mmgait}}).
	\item Our own dataset, featuring $8$ subjects (\mbox{\secref{sec:dataset}}). This dataset was collected from our own measurements, implementing the proposed system on an NVIDIA Jetson TX2\footnote{\url{https://developer.nvidia.com/embedded/jetson-tx2}} board paired with a Texas Instruments IWR1843BOOST \mbox{mm-wave} radar\footnote{\url{https://www.ti.com/tool/IWR1843BOOST}} operating in the \mbox{$77-81$~GHz} band.
\end{enumerate} 
 The Jetson board mounts an NVIDIA Tegra X2 GPU accelerator, the radar device is connected to it via USB and the communication is performed via UART ports, as shown in \fig{fig:jet-rad}. A camera was used to collect a video of the scene during the measurements and to label the dataset with the correct identities of the subjects. This setup poses some severe limitations on the amount of data that can be transferred in \mbox{real-time} to the NVIDIA processing device. Note that a more advanced solution such as an Ethernet connection would require additional hardware at an extra cost\footnote{\url{https://www.ti.com/tool/DCA1000EVM}}. 
The full system has been implemented in Python, using the TensorFlow library for the neural network classifiers. In \tab{tab:params}, the system parameters used in the evaluation are summarized. 

\subsection{Evaluation on the mmGait dataset}\label{sec:mmgait}

\begin{table}[t!] 
	\begin{center}
		\begin{tabular}{cc|cccc}
			\toprule	
			\multicolumn{2}{c}{{\bf Setup}}&\multicolumn{2}{c}{{\bf TCPCN (ours)}}& \multicolumn{2}{c}{{\bf mmGaitNet \cite{meng2020gait}}}\\
			\cmidrule(lr){3-4}\cmidrule(lr){5-6}
			Room &\# subj.&\multicolumn{1}{c}{linear}& \multicolumn{1}{c}{ free}&\multicolumn{1}{c}{linear}& \multicolumn{1}{c}{free} \\
			\cmidrule(lr){1-2}\cmidrule(lr){3-3}\cmidrule(lr){4-4}\cmidrule(lr){5-5}\cmidrule(lr){6-6}
			\texttt{room\_1}&$10$&$\boldsymbol{92.07}$&$\boldsymbol{70.31}$&$90.0$&$45.0$\\
			\texttt{room\_1}&$15$&$86.21$&$68.36$&$-$&$-$\\
			\texttt{room\_1}&$20$&$\boldsymbol{83.37}$&$63.97$&$80.0$&$-$\\
			\texttt{room\_2}&$30/29$&$89.34$&$64.73$&$-$&$-$\\
			\bottomrule
		\end{tabular} 		
	\end{center}
	\caption{Evaluation results on the mmGait dataset \mbox{\cite{meng2020gait}}. We report the  accuracy (\%) obtained by mmGaitNet according to the original paper \mbox{\cite{meng2020gait}} and the accuracy of our TCPCN, highlighting the best performance with a \textbf{bold} font. In the table, two columns show the results for linear and unconstrained motion. The dataset contains $29$ subjects for \texttt{room\_2} in the free motion case, and $30$ in the linear motion case. The symbol ``$-$'' is used for those cases for which no accuracy value is provided in~\mbox{\cite{meng2020gait}}. \label{tab:acc-mmgait}}
\end{table}

To assess the capabilities of TCPCN to effectively extract human gait features from point-cloud sequences, we test it on the publicly available mmGait dataset \mbox{\cite{meng2020gait}}, which contains measurements from two different evaluation rooms, \texttt{room\_1} and \texttt{room\_2}, including respectively $23$ and $31$ different subjects. The dataset contains sequences where subjects are constrained to walk along straight lines in front of the radar, and other sequences where they walk freely.
	
Next, we present a comparison between our neural network classifier, TCPCN, and the CNN proposed by the authors of \mbox{mmGait}, denoted by \mbox{mmGaitNet} \mbox{\cite{meng2020gait}}. The accuracy results obtained by TCPCN on a superset of the tests conducted by the authors in  \mbox{\cite{meng2020gait}} are shown in \mbox{\tab{tab:acc-mmgait}}. We stress that, for the sake of a fair comparison, for these results we just compared TCPCN with the CNN of \mbox{\cite{meng2020gait}}, without using our algorithms \mbox{\alg{alg:labeling}} and \mbox{\alg{alg:correction}}, as they would provide an additional performance increase.

For the results in \mbox{\tab{tab:acc-mmgait}}, we consider the mmGait traces recorded by a single TI IWR6843\footnote{\url{https://www.ti.com/tool/IWR6843ISK}} radar working in the $60-64$ GHz frequency band. The measurements for each subject are split according to a $80\%-20\%$ proportion to obtain training and test sets, as done in \mbox{\cite{meng2020gait}}.

TCPCN outperforms mmGaitNet in all the considered cases. The gap is particularly large in case the subjects walk freely: in this case, mmGaitNet reaches an accuracy of $45\%$ on $10$ subjects, as compared to an accuracy of $70.31\%$ for TCPCN. This difference is due to the high variety of patterns that occur in the presence of unconstrained motion. TCPCN is more robust to such variability thanks to its invariance to the ordering of the points in the data cloud.
The obtained performance on $30$ subjects is encouraging, leading to identification accuracies as high as $89.34\%$ and $64.73\%$ for linear and unconstrained motion, respectively. This shows that gait-based identification systems employing mm-wave radar sensors hold the potential of scaling to scenarios where the number of users is in the order of a few tens. 
Finally, we point out that the accuracy with $30$ subjects being higher than that with $15$ and $20$ is probably due to the fact that \texttt{room\_2} contains subjects who are more easily distinguishable than those from \texttt{room\_1}.

\subsection{Proposed dataset description}\label{sec:dataset}

\begin{table}[t!] 
	\begin{center}
		\begin{tabular}{lcr}
			\toprule
			\multicolumn{3}{c}{{\bf System parameters }} \\
			\midrule
			Antenna el. spacing & $d$ & $1.948$~mm\\
			Number of TX antennas & $N_{\mathrm{TX}}$ & $3$ \\ 
			Number of RX antennas & $N_{\mathrm{RX}}$ & $4$ \\ 
			Start frequency & $f_o$		& $77$~GHz      \\
			Chirp bandwidth & $B$	& $3.072$~GHz     \\
			Chirp duration & $T$       		& $60$~$\mu$s   \\
			Chirp repetition time  & $T_{\mathrm{rep}}$ &  $68$~$\mu$s\\
			No. samples per chirp & $M$    & $256$ \\
			No. chirps per seq. & $L$  & $64$ \\
			Frame rate & $1/\Delta t$       & $14.92$~fps    \\
			ADC sampling frequency & $1/T_f$   & $5$~MHz       \\
			Range resolution & $\Delta \tilde{R}$    & $4.88$~cm       \\
			Velocity resolution & $\Delta \tilde{v}$ & $14.9$~cm/s      \\
			\midrule 
			DBSCAN radius & $\varepsilon$ &   $0.4$~m   \\
			DBSCAN min. cluster dim. & $m_{\mathrm{pts}}$ &   $10$   \\
			\midrule
			Meas. range std & $\sigma_{R}$ &   $0.03$~m   \\
			Meas. az. angle std & $\sigma_{\theta}$ &   $\pi/24$~rad   \\
			Meas. ext. std & $\sigma_{\tilde{\ell}}, \sigma_{\tilde{w}}$ &   $0.05$~m   \\
			Meas. orient. std & $\sigma_{\tilde{\xi}}$ &   $\pi/6$~m   \\
			Process noise std & $\sigma_{a}$ &   $8$~m/s$^2$   \\ 
			Process ext. std & $\sigma_{\ell}, \sigma_{w}$ &   $0.001$~m   \\
			Process orient. std & $\sigma_{\xi}$ &   $\pi/24$~m   \\
			CJPDA bias term & $\beta$ &   $0.01$   \\
			$\texttt{m}/\texttt{n}$ logic parameters & $\texttt{m}/\texttt{n}$ &   $10/30$   \\
			\midrule
			Max point-cloud dim. & $n_{\max}$ &   $100$   \\
			Input time-steps & $K$ &   $30$   \\
			Moving avg. parameter & $\rho$ &   $0.99$   \\
			Decay parameter & $\gamma$ &   $0.999$   \\
			\midrule
			Dropout probability &$p_{\mathrm{drop}}$& $0.5$\\
			Regularization parameter&$\lambda_{L_2}$& $10^{-4}$\\
			Learning rate&$\eta$&$10^{-4}$\\
			\bottomrule
		\end{tabular} 		
	\end{center}
	\caption{Summary of the parameters of the proposed system. \label{tab:params}}
\end{table}

\begin{table}[t!] 
	\begin{center}
		\begin{tabular}{ccccccc}
			
			\toprule
			Subject & Age & Height [m] & Sex & $\ell$ [cm] & $w$ [cm] &  Frames \\
			\midrule
			$0$ & $26$ & $1.63$ &F&$43$&$22$&  $33,339$ \\ 
			$1$ & $26$ & $1.76$ & M&$52$& $23$& $33,514$ \\
			$2$ & $25$ & $1.85$ &M& $52$ & $24$ & $36,126$\\
			$3$ & $26$ & $1.72$ &M& $46$ & $16$ & $18,668$\\
			$4$ & $28$ & $1.69$ &M& $45$ & $22$ & $19,035$\\
			$5$ & $25$ & $1.61$ &F& $43$ & $20$ & $27,674$\\
			$6$ & $63$ & $1.77$ &M& $50$ & $24$ & $22,039$\\
			$7$ & $63$ & $1.58$ &F& $41$ & $20$ & $16,925$\\
			\bottomrule
		\end{tabular} 		
	\end{center}
	\caption{Details on the subjects involved in the measurements. \label{tab:subjs}}
\end{table}

To further validate the proposed system, we built our own dataset using four different rooms: three to collect training data and one for testing purposes. This arrangement of data and rooms was intentionally adopted to asses the generalization capabilities of the proposed system. Eight subjects were involved in the measurements, see \tab{tab:subjs}. 
\begin{figure*}[t!]
	\begin{center}   
		\centering
		\subcaptionbox{Estimated trajectories and extensions of three subjects walking freely in the test room. \label{fig:tracks}}[5.7cm]{\includegraphics[width=5.4cm]{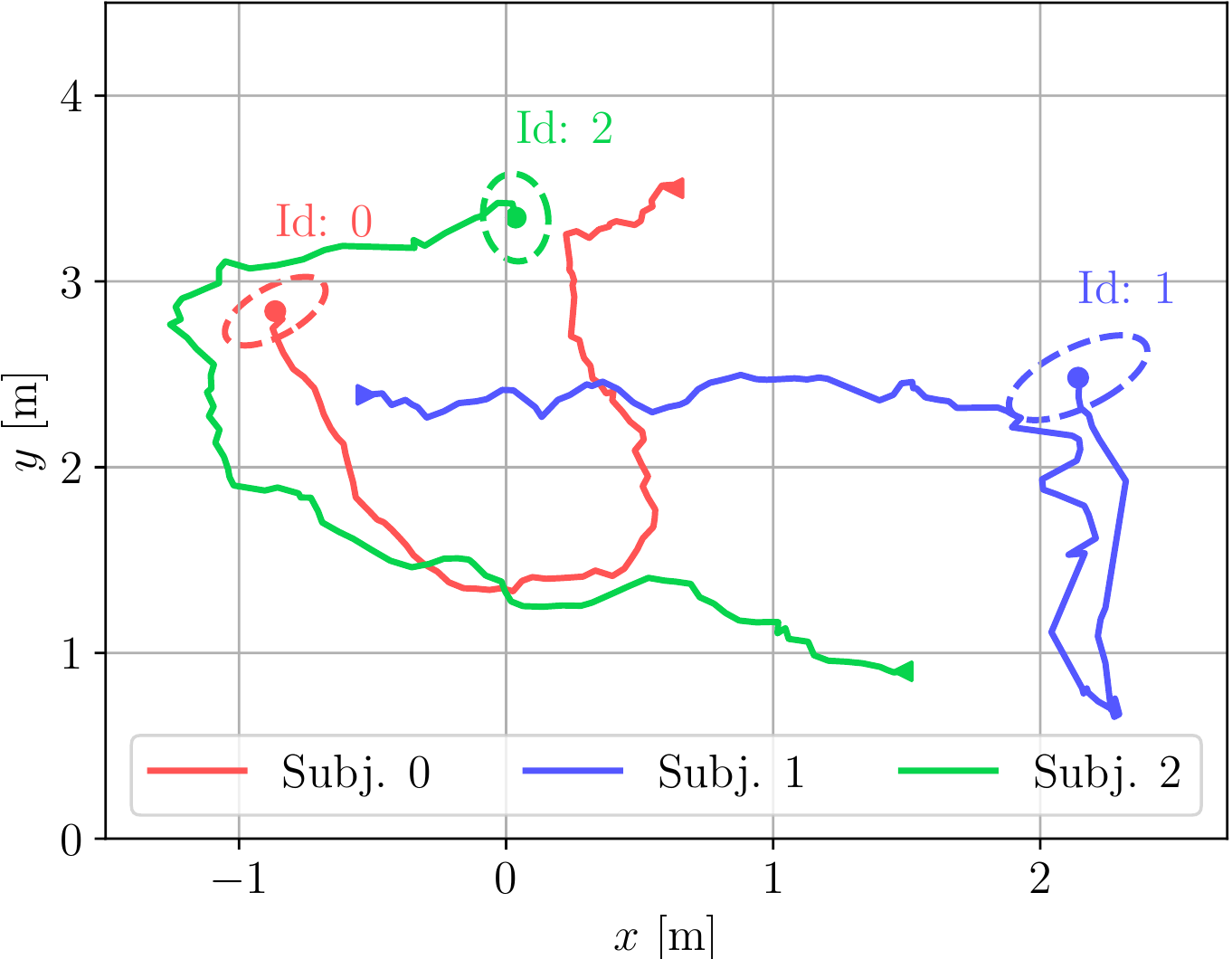}}
		\subcaptionbox{Evolution of the extension estimates for the three subjects.\label{fig:ellipses}}[5.7cm]{\includegraphics[width=5.4cm]{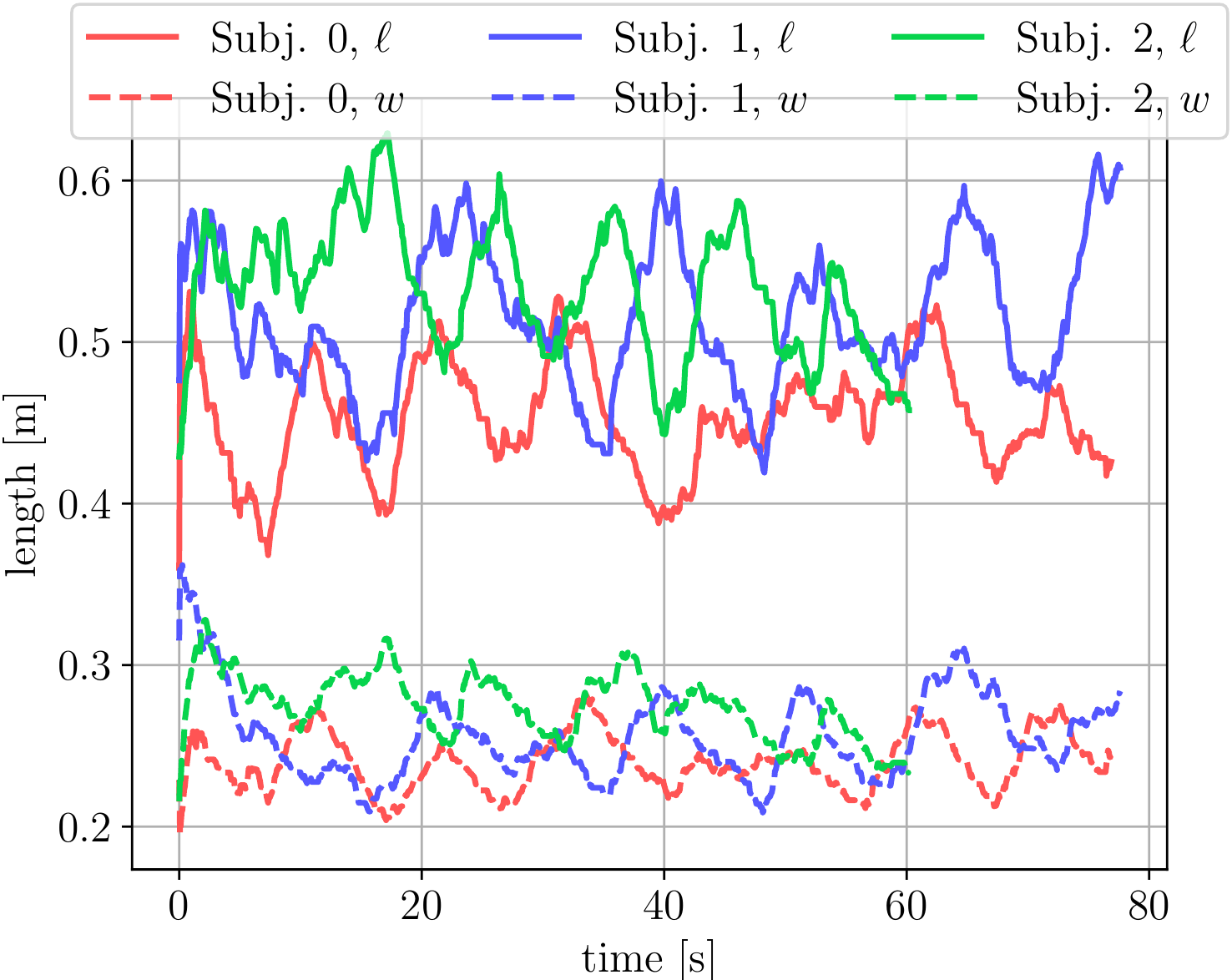}}
		\subcaptionbox{Average MOTA using tracking only or joint tracking and identification.\label{fig:tot}}[5.7cm]{\includegraphics[width=5.6cm]{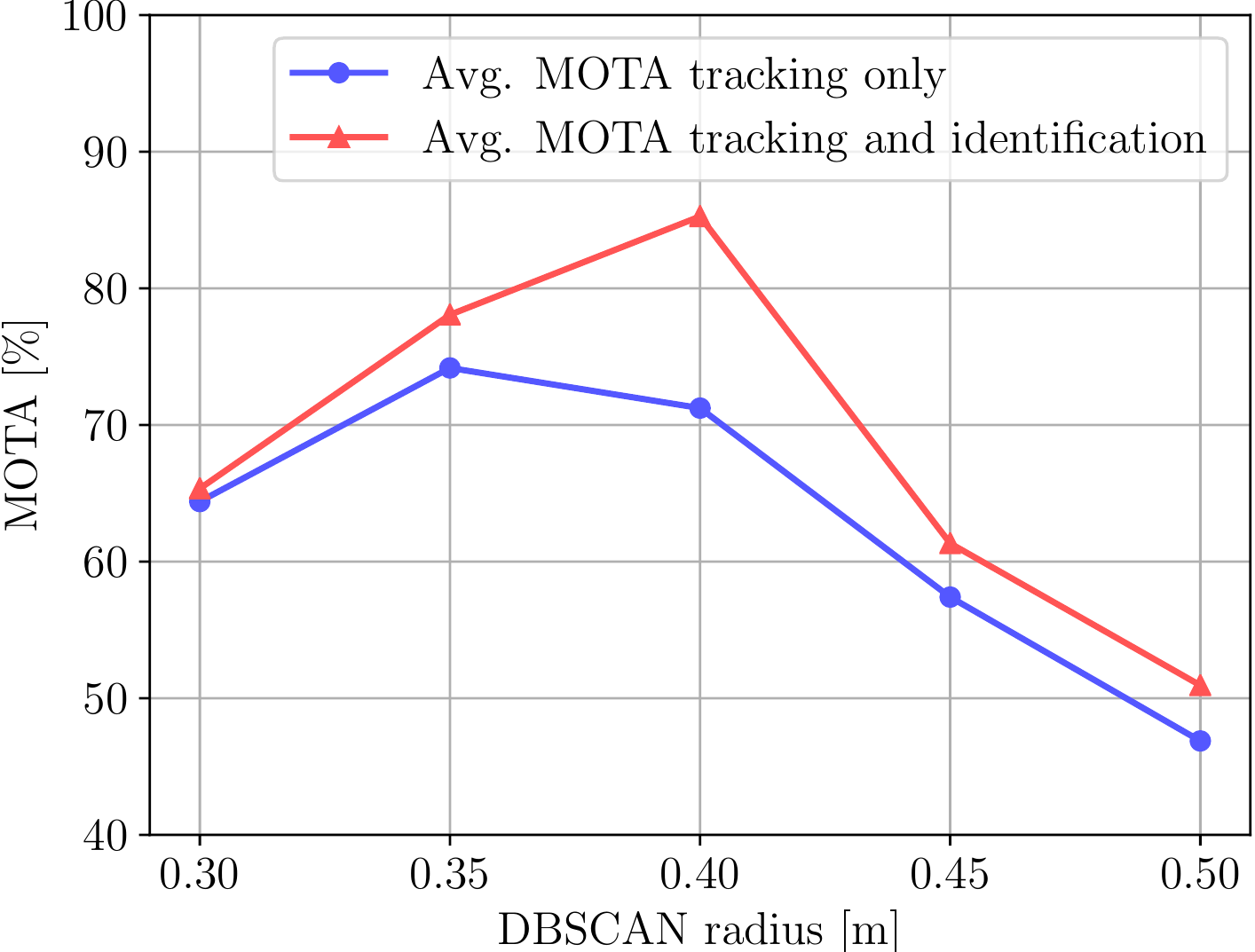}}
		\caption{Tracking system evaluation.}
		\label{fig:tracking-res}
	\end{center}
\end{figure*}

\textbf{Training:} the training rooms are two research laboratories, of size $8 \times 8$ meters and $8 \times 3$ meters, respectively, containing desks, furniture and technical equipment, and a furnished living room of size $8 \times 5$ meters. In the first room, due to space limitations, the area used for the training measurements is a rectangular space of size $3 \times 5$ meters. To collect the training data, one subject at a time walked freely for an amount of time ranging from $1$ to $5$ minutes.
Note that, in all our measurements the subjects are allowed to cover a distance of up to $6$~m from the radar, within its field-of-view of $\pm 60^{\circ}$.

The measurement campaign was repeated across different days, acquiring from $20$ to $40$ minutes of data per subject. Taking into account different days, we aimed at reducing the effect of clothing or daily patterns in the way of walking. Prior to the actual training phase, the \mbox{point-clouds} data were \mbox{pre-processed} as described in \secref{sec:pc-preproc} and grouped into sequences of $K=30$ consecutive frames, leaving an overlap of $20$ frames between different sequences. To reduce overfitting, we artificially augmented the training data by applying random shuffling of the points in each \mbox{point-cloud} and adding random noise to each point, drawn from a uniform distribution in the interval $[-0.1, 0.1]$. To select the neural network hyperparameters, a portion of the training data (one sequence of approximately $2,250$ frames per target) was used as a validation set.

\textbf{Test:} the test room is a $7 \times 4$ meters research laboratory, whose measurement area is free of furniture (see \fig{fig:setup}). 
We stress that, while training is performed on up to $8$ {\it single subjects}, all our test sequences include multiple targets concurrently moving in the test environment. This leads to blockage events, i.e., when a subject occludes the \mbox{line-of-sight} (LoS) between the radar and another target, resulting in bursts of frames where the blocked subject goes undetected.

The measurement sequences contained in the test dataset are split as follows:
\begin{enumerate}
	\item $10$ sequences of $80$ seconds ($1,200$ frames) with $3$ subjects. These are further split into $5$ sequences where the subjects were constrained to walk following a linear movement at their preferred speed (back and forth across predefined linear paths), and $5$ sequences where they could walk freely, following any trajectory in the available space, as shown in \fig{fig:subjects}. This leads to unpredictable trajectories that can cover the whole field-of-view of the radar sensor ($\pm 60^{\circ}$) and distances up to $6$~m.
	Moreover, in all our experiments user trajectories intersect frequently, leading to ambiguities in the data association, and making tracking more challenging.
	\item $10$ sequences of $80$ seconds with $2$ subjects, split into $5$ sequences with a linear walking movement, and $5$ sequences where the subjects walk freely.
\end{enumerate}

\subsection{Tracking phase evaluation}\label{sec:track-ph}

\begin{figure*}[t!]
	\begin{center}   
		\centering
		\subcaptionbox{ \label{fig:649}}[5.5cm]{\includegraphics[width=4.7cm]{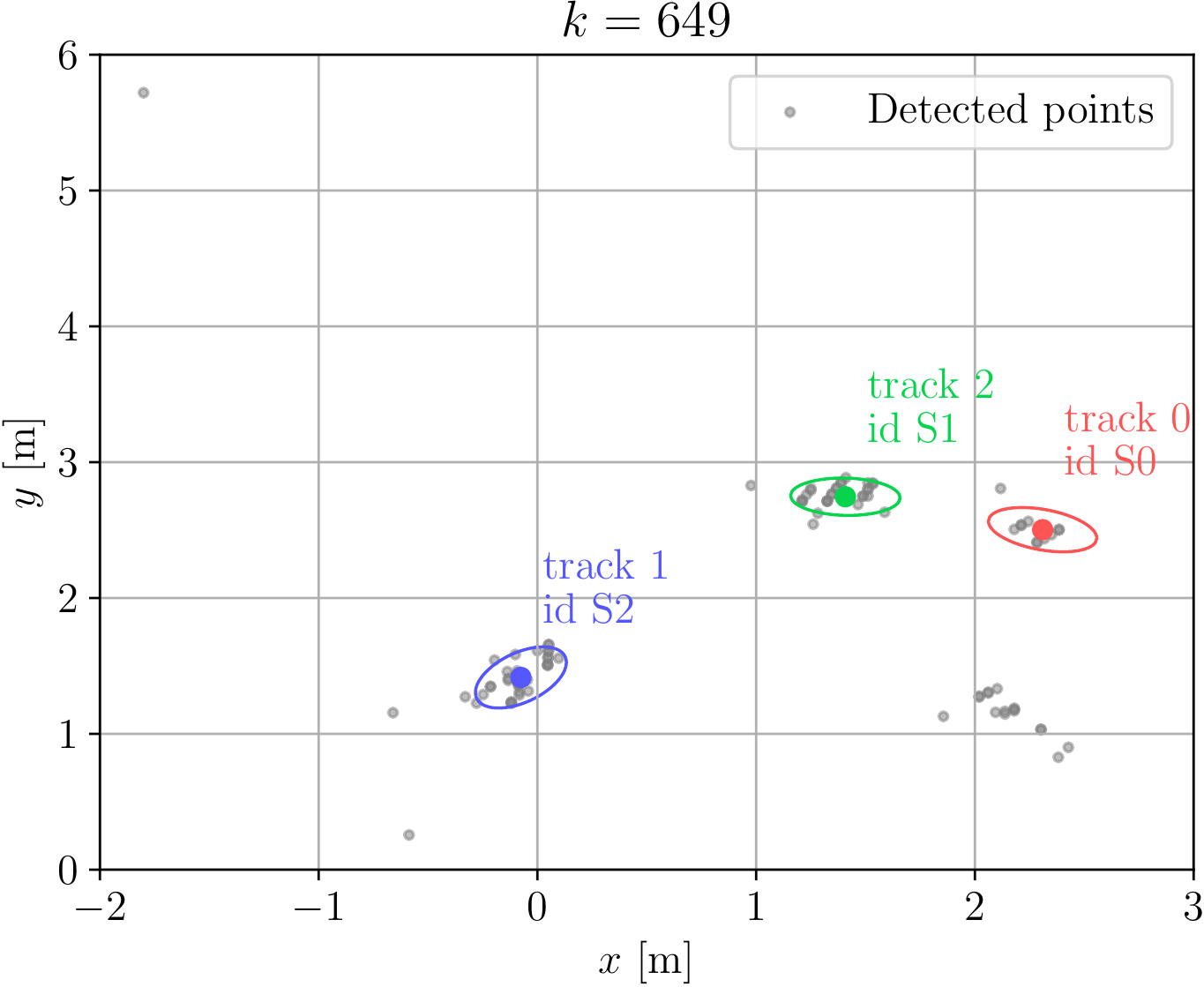}}
		\subcaptionbox{\label{fig:669}}[5.5cm]{\includegraphics[width=4.7cm]{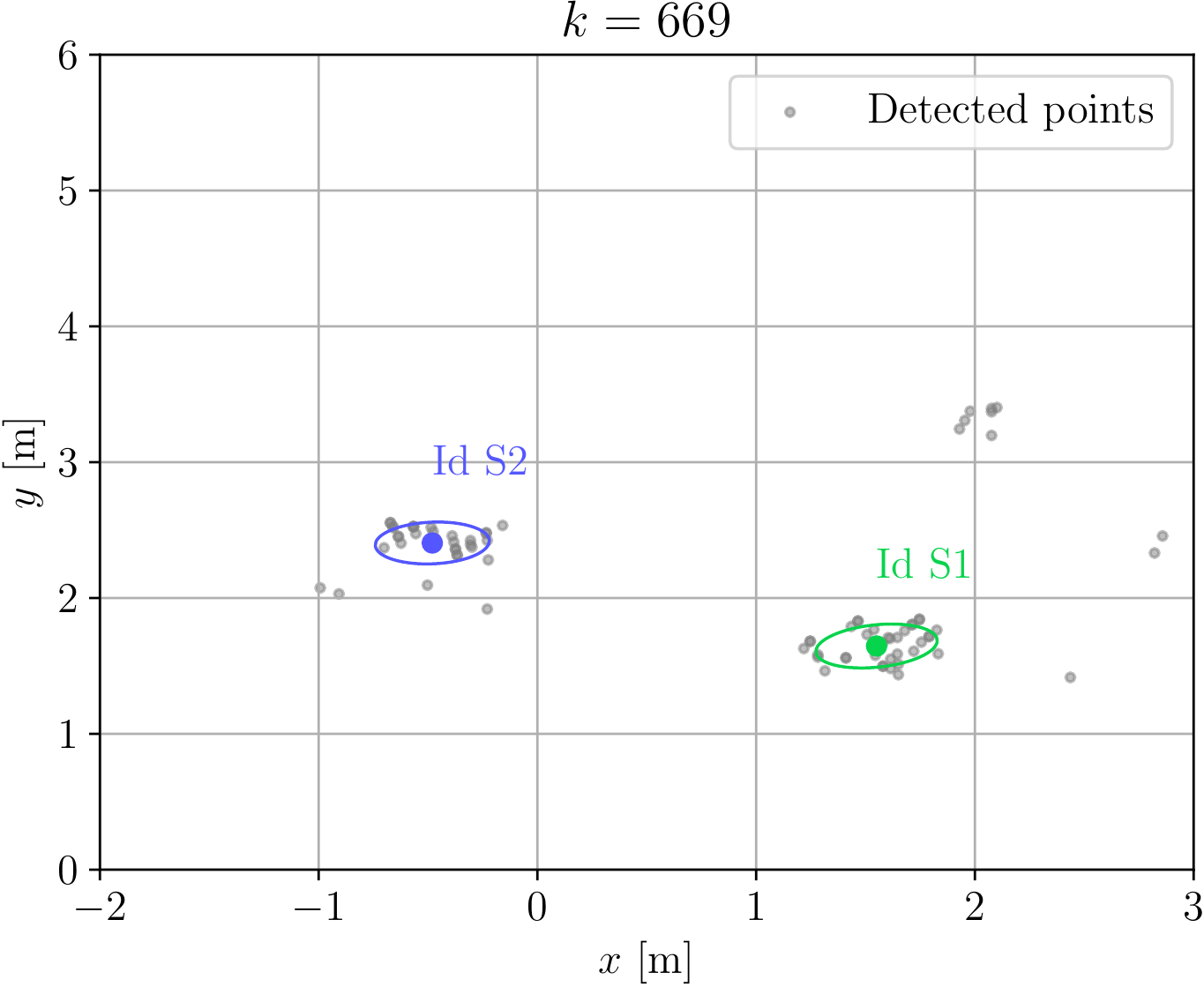}}
		\subcaptionbox{\label{fig:700}}[5.5cm]{\includegraphics[width=4.7cm]{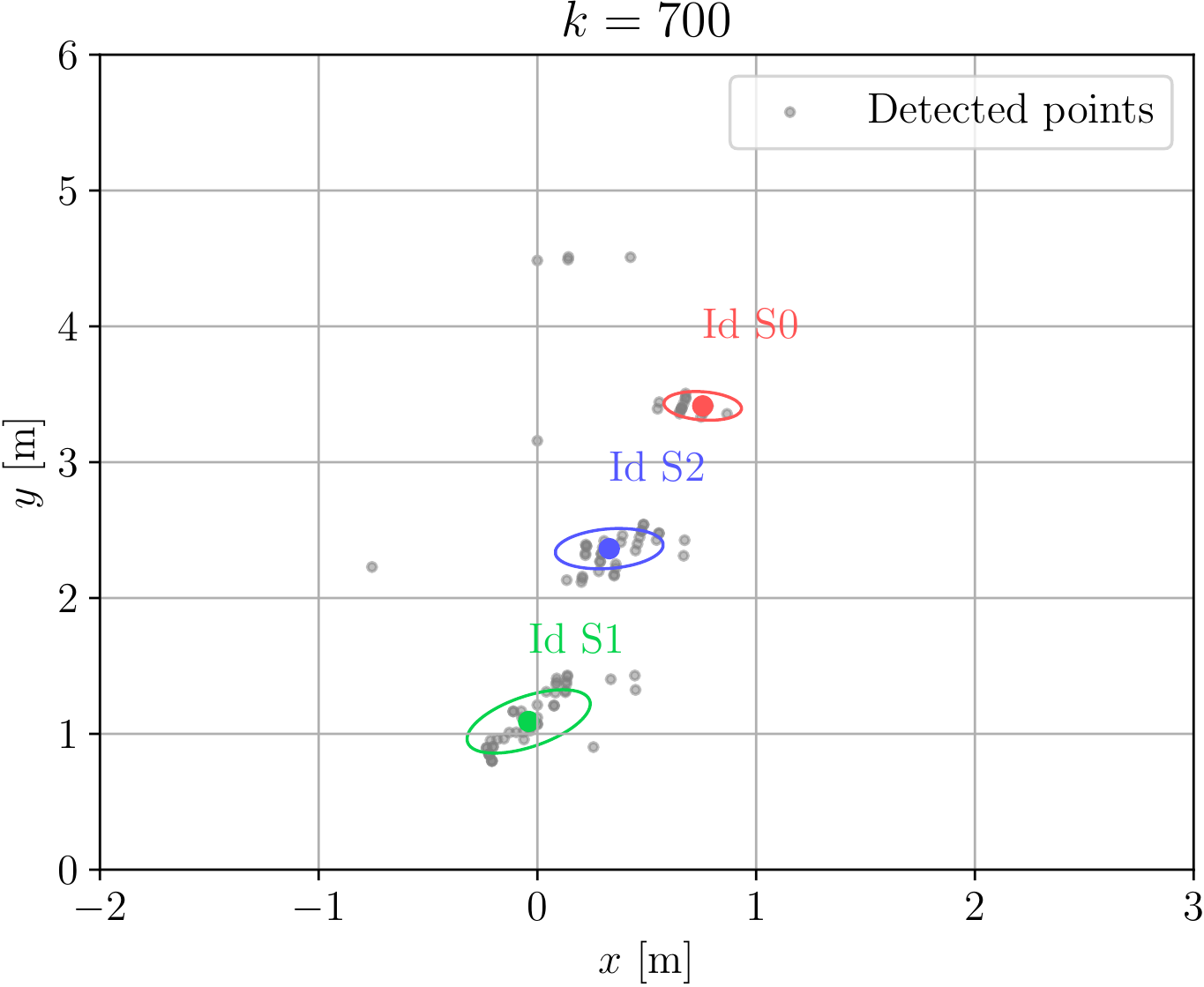}}
		\subcaptionbox{ \label{fig:649_tr}}[5.5cm]{\includegraphics[width=4.7cm]{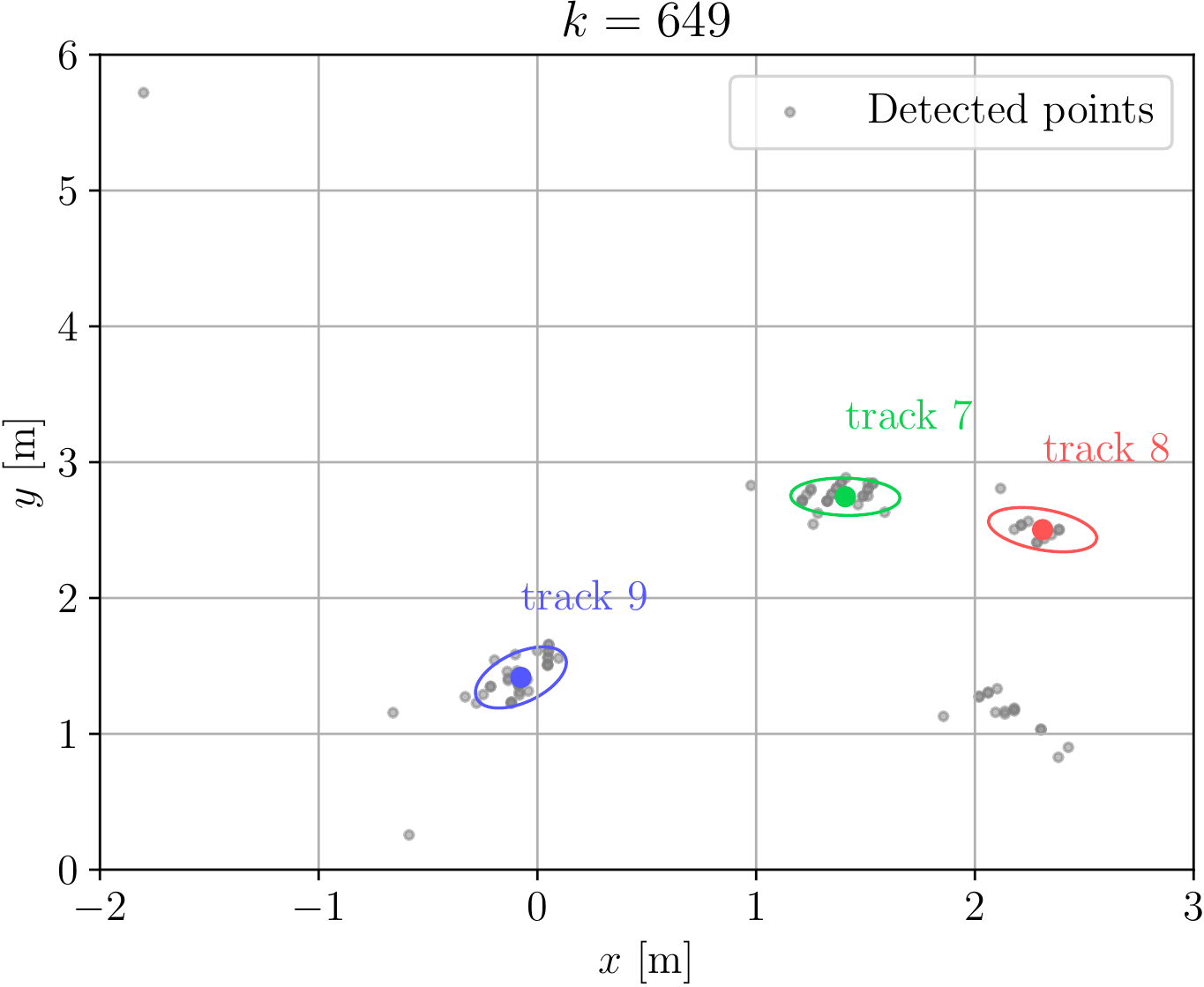}}
		\subcaptionbox{\label{fig:669_tr}}[5.5cm]{\includegraphics[width=4.7cm]{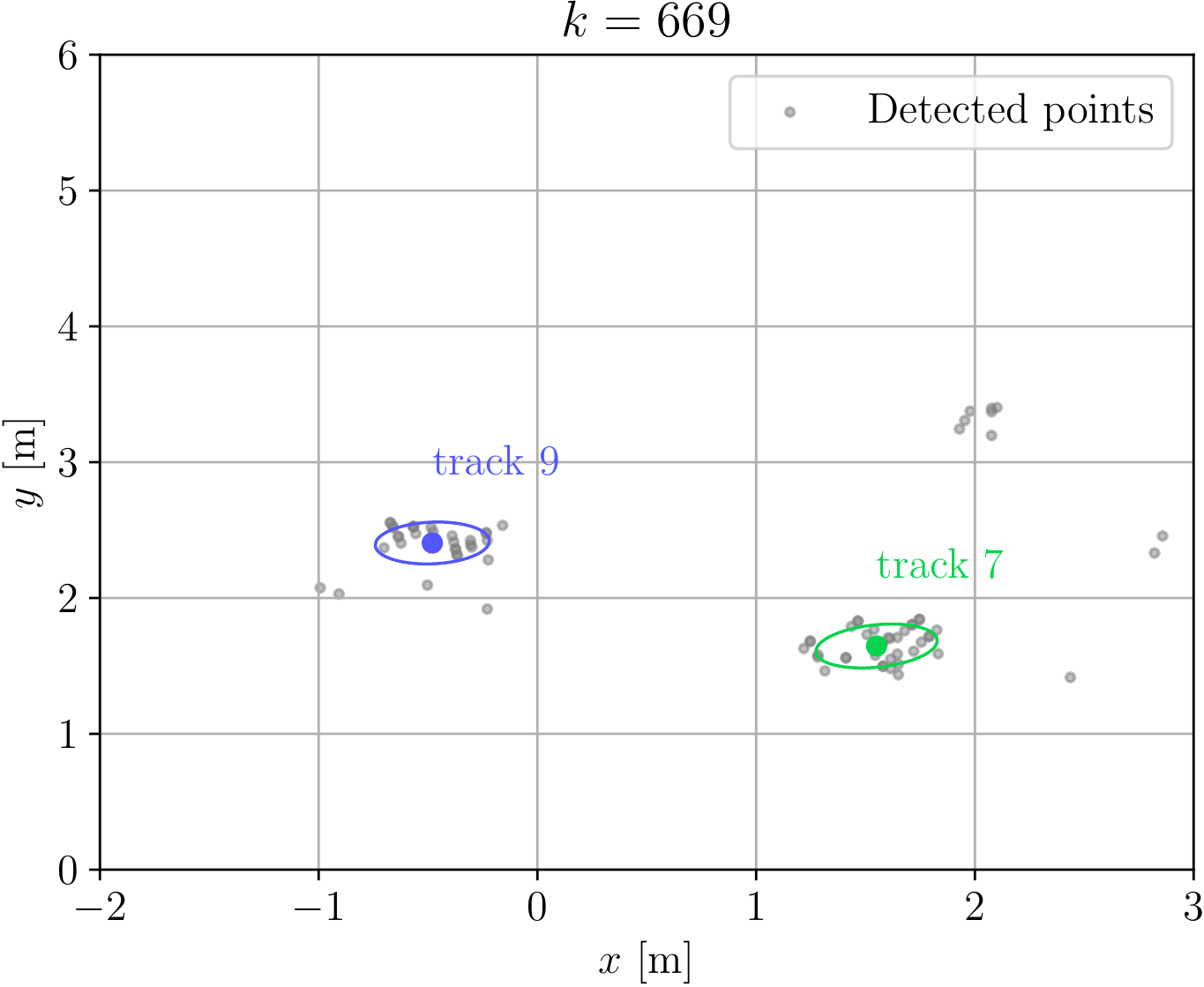}}
		\subcaptionbox{\label{fig:700_tr}}[5.5cm]{\includegraphics[width=4.7cm]{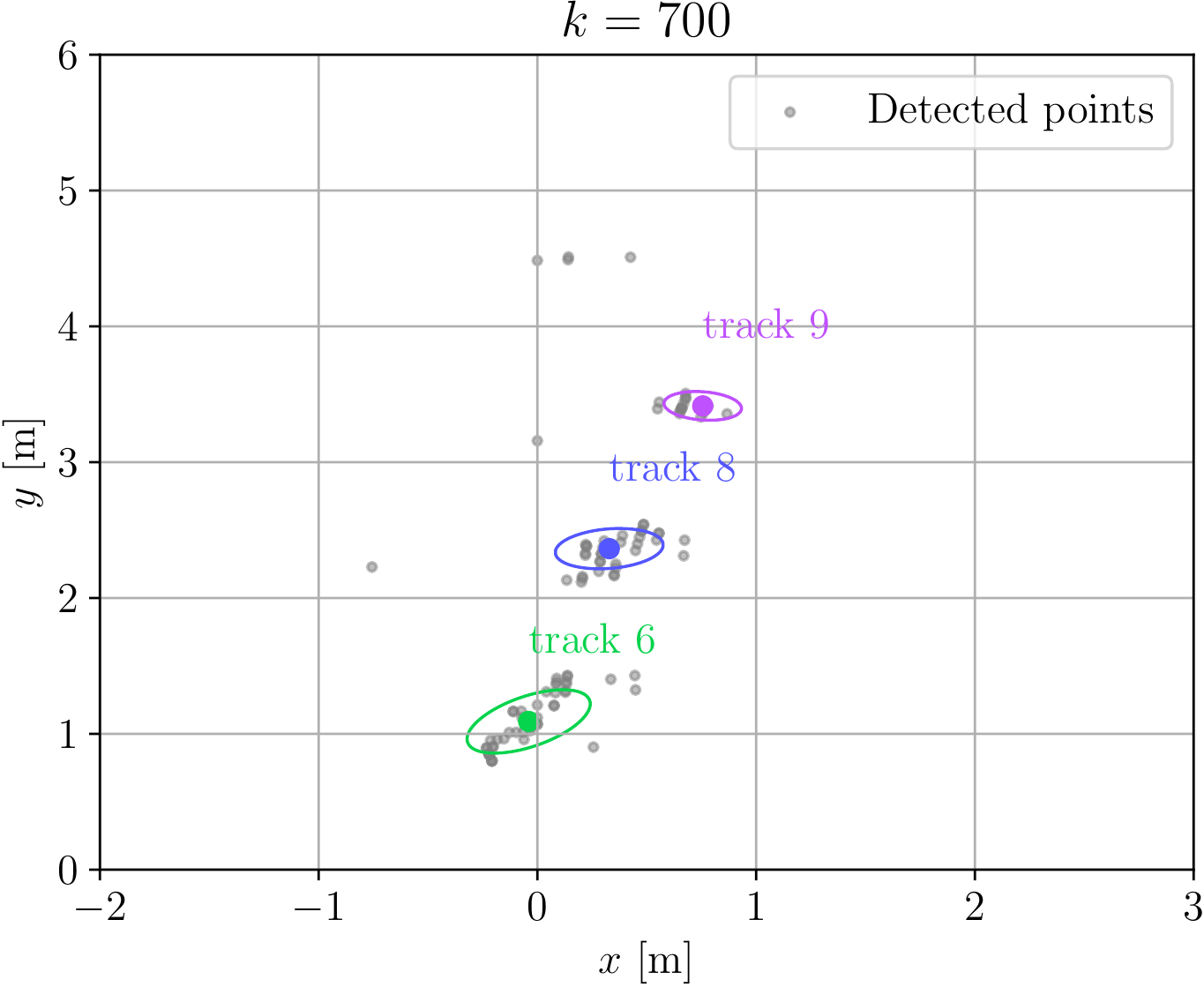}}
		\caption{Proposed identification algorithm (a - b - c) compared to a standalone tracking approach (d - e - f) on the $x - y$ plane. Subject $0$ (S$0$) is lost at time $k = 669$ and tracked again at time $k=700$. By joint use of tracking and identification algorithms, the new track $3$ is correctly \mbox{re-associated} with S$0$ (c), i.e., track $3$ is mapped back onto track $0$. Instead, the sole use of tracking would lead to the initialization of a new track for the same subject (f), causing a mismatch.}
		\label{fig:id-alg-succ}
	\end{center}
\end{figure*}

In \fig{fig:tracks}, we show example trajectories followed by the three targets in one of the test sequences. In this experiment, the \mbox{CM-KF} succeded in identifying and reconstructing the trajectory of each target, even in the presence of complex and strongly \mbox{non-linear} movement. The \mbox{NN-CJPDA} data association logic was found to be very robust, as long as the targets are correctly separated by DBSCAN into disjoint \mbox{point-cloud} clusters.

In all the test measurements the main difficulty faced by the system was that of handling blockage events that span over a large number of frames, e.g., more than \mbox{$2-3$} second long. The number of such events increases significantly when more subjects are added to the monitored environment.
We empirically assessed that, using a single radar sensor with the resolution and communication capabilities considered in this work, going beyond three freely moving subjects at a time in such a small indoor environment leads to insufficient tracking and identification accuracy due to blockage. This is coherent with the findings in the literature, e.g.~\cite{meng2020gait}, where two radars placed in different locations were used to compensate for these facts.

\fig{fig:ellipses} shows the results of the extension estimation across a full test sequence for all subjects. The expected shape enclosing a human target is correctly estimated: the ellipse axes are coherent with typical shoulder widths, $\ell$, and thorax widths along the sagittal plane, $w$. The estimated value varies depending on the position of the target with respect to the radar: this is due to the fact that the received \mbox{point-clouds} contain a smaller number of points as the distance increases, due to propagation losses. Despite this fact, the average values are still proportional to the true subjects' extensions, as it can be checked by comparing \fig{fig:ellipses} with \tab{tab:subjs}.
\begin{table*}[t!] 
	\begin{center}
		\begin{tabular}{cccccccccc}
			\toprule
			&\multicolumn{3}{c}{{\bf Test 2 sub. train 3}}& \multicolumn{3}{c}{{\bf Test 3 sub. train 3}} & \multicolumn{3}{c}{{\bf Test 3 sub. train 8}}\\
			\cmidrule(lr){2-4}\cmidrule(lr){5-7}\cmidrule(lr){8-10}
			&linear& \multicolumn{2}{c}{free}&linear& \multicolumn{2}{c}{free}&linear& \multicolumn{2}{c}{free} \\
			\cmidrule(lr){2-2}\cmidrule(lr){3-4}\cmidrule(lr){5-5}\cmidrule(lr){6-7}\cmidrule(lr){8-8}\cmidrule(lr){9-10}
			[\%]& Id. acc.& Id. acc.&MOTA& Id. acc.&Id. acc.&MOTA&Id. acc.&Id. acc.&MOTA\\
			\midrule
			Seq. 1&$98.67$&$99.61$&$98.71$&$100$&$99.67$&$99.06$&$96.95$&$92.35$&$99.06$\\
			Seq. 2&$100$&$99.75$&$98.71$&$100$&$99.91$&$84.14$&$99.81$&$96.17$&$84.14$\\
			Seq. 3&$95.26$&$96.91$&$86.42$&$100$& $91.79$&$94.11$&$100$&$88.14$&$90.19$\\
			Seq. 4&$99.54$&$100$&$99.62$&$90.43$&$100$&$76.36$&$90.43$&$89.46$&$76.36$\\
			Seq. 5&$99.34$&$100$&$97.96$&$99.37$&$92.44$&$72.61$&$97.04$&$92.02$&$72.61$\\
			\midrule
			\textbf{Average} &$\boldsymbol{98.56}$&$\boldsymbol{98.98}$&$\boldsymbol{96.28}$&$\boldsymbol{97.96}$&$\boldsymbol{96.76}$&$\boldsymbol{85.26}$&$\boldsymbol{96.85}$&$\boldsymbol{91.62}$&$\boldsymbol{84.47}$\\
			\bottomrule
		\end{tabular} 		
	\end{center}
	\caption{Accuracy and MOTA obtained with $2$ and $3$ subjects moving in the test room. We report the results both when the subjects follow linear trajectories (``linear'') and when they move freely (``free''). With ``Test $x$ sub. train $y$'' we denote the fact that the TCPCN used for the identification was trained on the single-target measurements of $y$ subjects and tested on multi-target sequences containing $x$ subjects simultaneously. \label{tab:acc-res}}
\end{table*}

To evaluate the capability of the proposed system towards tracking human subjects and the improvement brought by combining tracking and identification algorithms, we use the popular multiple object tracking accuracy (MOTA) metric~\cite{bernardin2006multiple}. The MOTA conveniently summarizes the ratio of missed targets ($\mathrm{miss}$), false positives ($\mathrm{fp}$) and track mismatches ($\mathrm{mm}$), over the number of ground truth targets ($\mathrm{gt})$ in each time frame $k$ of the test sequence, formally,
\begin{equation}
	\label{eq:MOTA}
	\mathrm{MOTA} = 1 - \frac{\sum_k \left(\mathrm{miss}_k + \mathrm{fp}_k + \mathrm{mm}_k\right)}{\sum_k \mathrm{gt}_k}.
\end{equation} 
The value of $\mathrm{gt}_k$ was obtained from a reference video, as mentioned at the beginning of \mbox{\secref{sec:results}}.

In \mbox{\fig{fig:tot}}, we show the MOTA obtained for different values of the DBSCAN radius, $\varepsilon$, for the NN-JPDA algorithm and our method, where NN-JPDA is used in conjunction with \mbox{\alg{alg:labeling}} and \mbox{\alg{alg:correction}} (subject identification and label correction). Note that, with the standard NN-JPDA tracking algorithm, when a track is deleted and \mbox{re-initialized}, it is counted as a mismatch in \mbox{\eq{eq:MOTA}}, significantly lowering the MOTA. Moreover, data association errors can lead to track swaps when the trajectories of two subjects intersect. The MOTA obtained in this case is plotted as a blue curve in \mbox{\fig{fig:tot}}. The red curve instead represents the improved MOTA, obtained by \textit{(i)} merging together all the tracks associated with the same subject's identity, as described in \mbox{\secref{sec:id-alg}}, and \textit{(ii)} correcting track swaps using \mbox{\alg{alg:correction}}. For the sake of clarity, in \mbox{\fig{fig:id-alg-succ}} we exemplify \mbox{step \textit{(i)}}, which significantly improves the results by mitigating the effect of losing and re-initializing tracks.
	
From \mbox{\fig{fig:tot}}, we see that for the optimal value $\varepsilon = 0.4$~m, the integration of tracking and identification provides an improvement of almost $20\%$ in terms of MOTA. Remarkably, this is obtained at almost no additional complexity, by just feeding back the identity information to the tracking block.


\subsection{Accuracy results}\label{sec:acc-res}


In \tab{tab:acc-res}, we report the person identification accuracy obtained with the proposed method on the test sequences described in \secref{sec:dataset}. For the unconstrained walks, we also report the corresponding MOTA. The \mbox{per-subject} identification accuracy is computed using the \mbox{time-steps} in which the subject is correctly tracked, and is defined as the fraction of \mbox{time-steps} where a subject, besides being tracked, is also correctly identified. The final accuracy on a test sequence is obtained by taking the average accuracy on each subject, weighted by the total number of frames in which he/she is detected and tracked by the system.

In our tests, the number of subjects used for training is set as either $3$ or $8$ to assess how the system performs with an increasing number of targets. In \mbox{\tab{tab:acc-res}}, this is indicated with ``Test $x$ sub. train $y$'', where $x$ and $y$ respectively refer to the number of subjects in the training set and those who are simultaneously present in the test data. The accuracy ranges from a maximum of $98.98\%$ down to $91.62\%$, with the latter achieved for the most challenging case where $3$ concurrent subjects have to be identified among a set of $8$.

Differently from the results on mmGait (see \mbox{\secref{sec:mmgait}}), there are no significant deviations in the identification performance between linear and unconstrained motion. This is due to the proposed identification algorithm, which lowers the effect of turns and \mbox{non-linear} movements that are likely to impact the classification accuracy. The MOTA is instead significantly lower with three targets, because of the more frequent blockage events (more misses and mismatches).

\begin{figure}[t!]
	\begin{center}   
		\includegraphics[width=7.5cm]{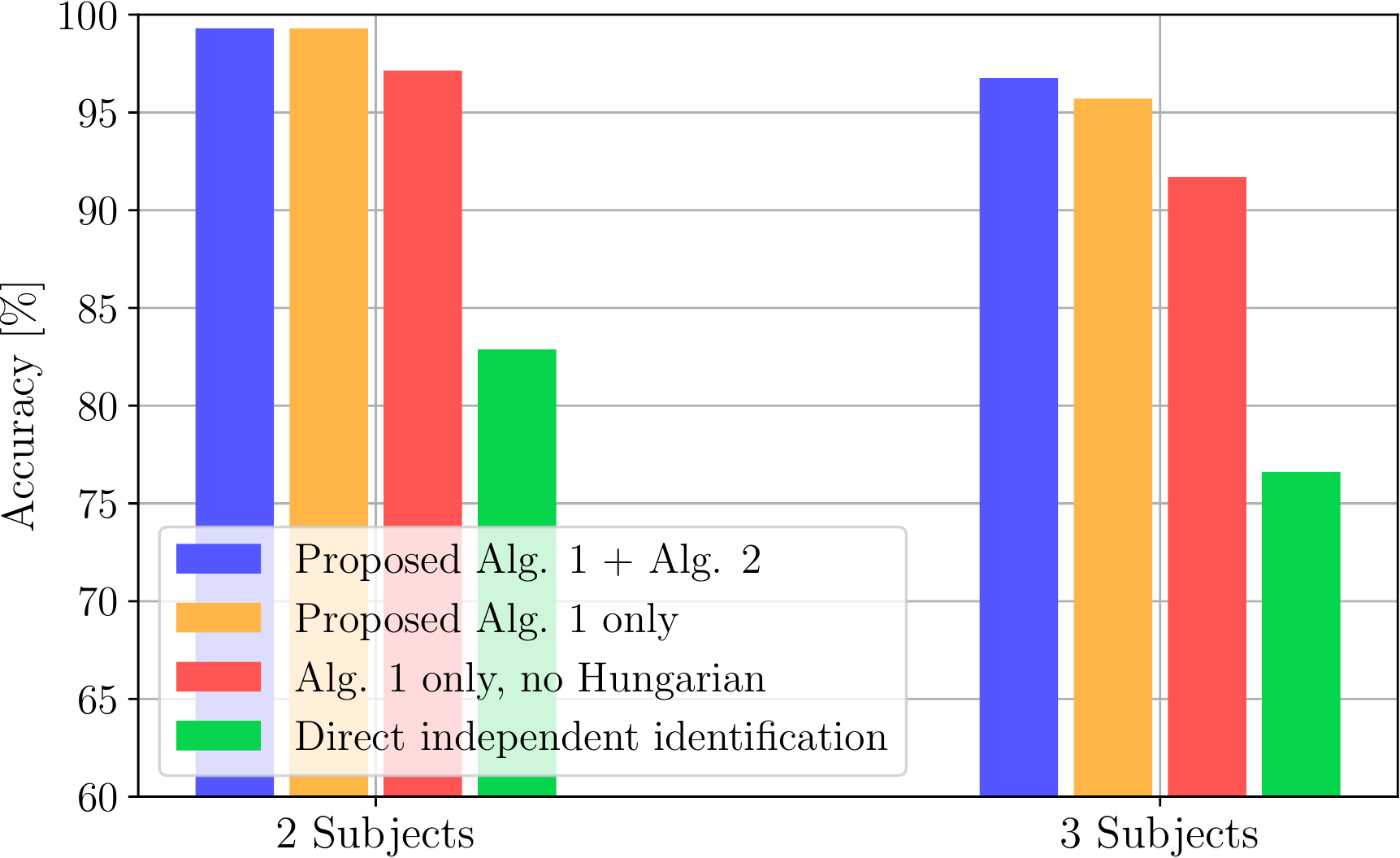} 
		\caption{Accuracy of the proposed identification algorithm.}
		\label{fig:id-algo}
	\end{center}
\end{figure}

\fig{fig:id-algo} shows the average accuracy obtained over the free-walking test sequences by \textit{(i)} using the proposed solution (\alg{alg:labeling} and \alg{alg:correction}), \textit{(ii)} using \alg{alg:labeling} only, \textit{(iii)} using \alg{alg:labeling} without the Hungarian method, and \textit{(iv)} identifying each subject at each time step $k$ by solely using the point-cloud data at time $k$, and estimating the identity as $\argmax\tilde{\boldsymbol{y}}_k^t$. For this evaluation the TCPCN was trained on $3$ subjects. Note that with $2$ subjects \textit{(i)} and \textit{(ii)} lead to about the same performance, but \alg{alg:correction} leads to a slight improvement with $3$ subjects, as tracking errors caused by track swaps due to blockage are more frequent in this case.

\subsection{Impact of temporal filtering parameters}\label{sec:train-params}

\begin{figure}[t!]
	\begin{center}   
		\includegraphics[width=7cm]{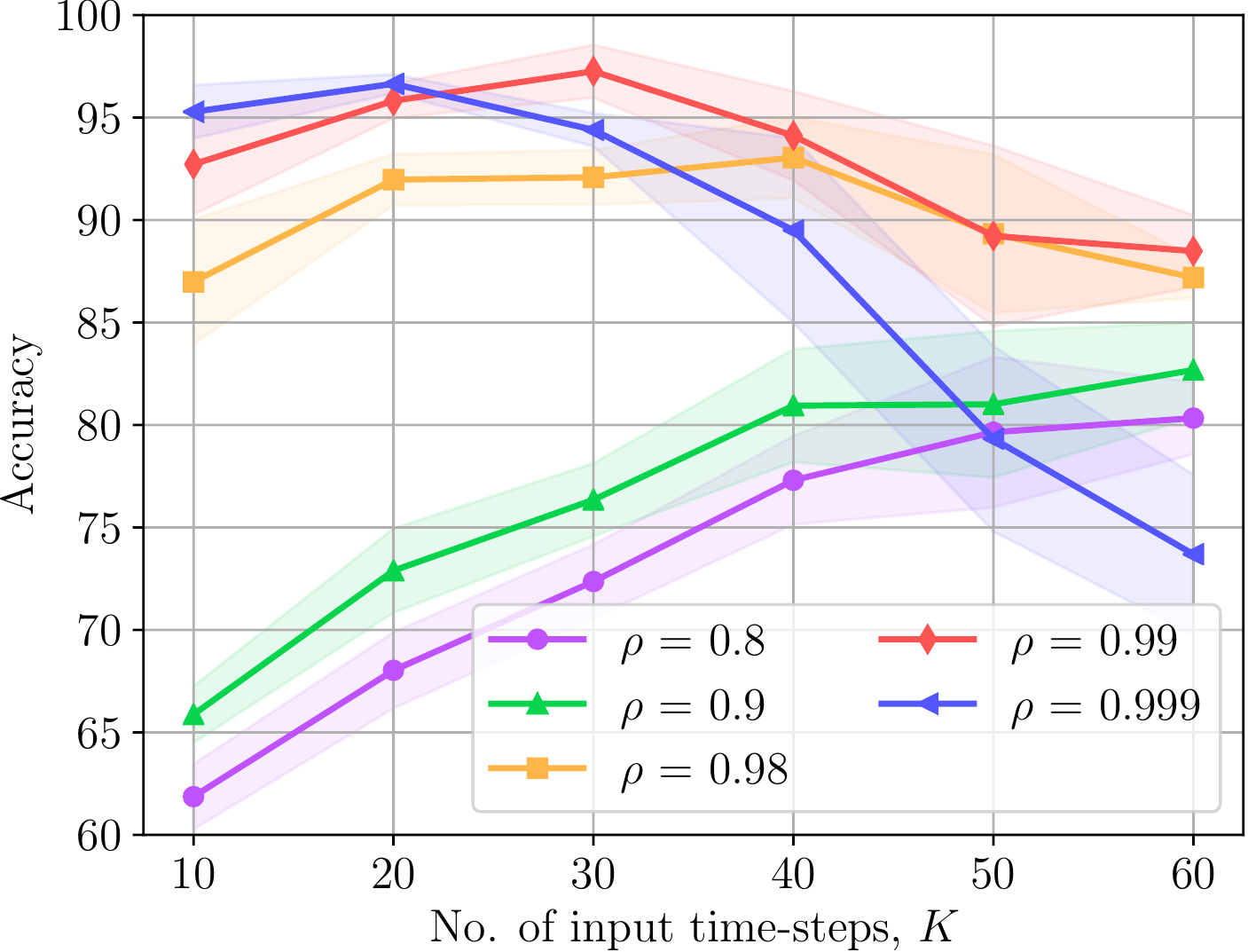} 
		\caption{Effect of varying $K$ and $\rho$ on the identification accuracy.}
		\label{fig:rho}
	\end{center}
\end{figure}

Now, we analyze the impact of $K$ and $\rho$, i.e., the number of input time steps and the moving average smoothing parameter, respectively. These parameters are intimately connected, as they both control the dependence of the current output on past frames. In \fig{fig:rho}, we show the average accuracy computed on $10$ different trainings of TCPCN with $Q=3$ subjects, when tested on $3$ subjects moving freely. The shaded areas represent $95$\% confidence intervals. In the abscissa, we  vary $K$, plotting a different curve for several selected values of $\rho$. Lower values of $\rho$, e.g., $0.8$ or $0.9$, lead to a lower performance, as the memory of the moving average filter in these cases is too short to introduce stability in the classification (it corresponds to $5$ and $10$ time steps for $\rho=0.8$ and $0.9$, respectively) and high values of $K$ are required to get an  accuracy beyond $80$\%. Increasing $\rho$ has the effect of moving the point of maximum accuracy towards lower values of $K$. From our results, we recommend using $K=30$ (two seconds of radar readings) and $\rho=0.99$, as these values lead to the best average accuracy while keeping the system sufficiently reactive, with a moving average memory of approximately $100$ time steps (between $6$ and $7$ seconds).

\subsection{Importance of point-cloud features}\label{sec:feat-imp}

In \mbox{\tab{tab:feat-comp}} we show the accuracy results of the sole TCPCN (no \mbox{\alg{alg:labeling}} and \mbox{\alg{alg:correction}}) considering $8$ single targets, by leaving out some of the point-cloud features in $\boldsymbol{p}_r$. Specifically, we trained and tested the NN by selectively leaving out the received power (no-$P$), the velocity (no-$v$), the $z$ coordinate (no-$z$) or the $x-y$ coordinates (no-$xy$). This evaluation provides insights on the importance of each of these features towards identifying the subjects. In particular, removing the velocity, $x-y$ or $z$ coordinates led to the largest reduction in accuracy, suggesting that these carry the most useful information. In addition, \mbox{\tab{tab:feat-comp}} proves that our method mostly relies on movement-related features rather than on the reflectivity of the target (related to the received power). We remark that this is key to gain robustness to reflectivity changes due to different clothing or other environmental factors, and the lower importance of certain features is enforced by the learning procedure, which has automatically learned it by processing data from the same subjects across different days (wearing different clothes, etc.) and environments.

\begin{table}[t!] 
	\begin{center}
		\begin{tabular}{cccccc}
			\toprule
			&$\boldsymbol{\mathrm{all}}$&$\boldsymbol{\mathrm{no-}P}$& $\boldsymbol{\mathrm{no-}v}$ & $\boldsymbol{\mathrm{no-}z}$&$\boldsymbol{\mathrm{no-}xy}$ \\
			\midrule
			Acc. [\%]&$82.08$ & $79.66$ & $65.89$ & $66.53$ & $65.52$ \\
			\bottomrule
		\end{tabular} 		
	\end{center}
	\caption{TCPCN accuracy (no \mbox{\alg{alg:labeling}} and \mbox{\alg{alg:correction}}) on $8$ single targets using: all the point-cloud features in $\boldsymbol{p}_r$ (all), selectively leaving out the received power information (no-$P$), the velocity (no-$v$), the $z$ coordinate (no-$z$) or the $x-y$ coordinates (no-$xy$). \label{tab:feat-comp}}
\end{table}

\subsection{Real-time implementation requirements}\label{sec:real-time}

Operating the proposed system in \mbox{real-time} poses constraints on the execution time of each processing block, and on the choice of the size and structure of the NN classifier. We measured the computation time needed by each block, respectively denoting by $t_p$ the time needed to run the point-cloud extraction module running on the radar device (including the chirp sequence transmission, three DFTs along the fast time, slow time and angular dimension and the CA-CFAR detector), by $t_c$ the time to transmit the data using the UART port, by $t_t$ the execution time of the DBSCAN clustering algorithm, the \mbox{CM-KF} tracking step and the data association, and by $t_i$ the inference time of the classifier. We found that while $t_p$ is stable and strictly lower than $10$~ms, $t_c$ is highly variable, mostly because of the variable number of detected points in the scene, and ranges between $0$~ms (when no points are detected) and $25$~ms (with $3$ subjects). The clustering and tracking take on average $t_t=12$~ms with $3$ subjects, with very low variance.
Being the radar frame duration $\Delta t \approx 67$~ms, the identification step has meet the inequality \mbox{$t_i < \Delta t - \max t_p - \max t_c - t_t \approx 20$~ms}. In the next section, we present a comparison between the proposed approach and two works from the literature in terms of accuracy and inference time, taking these considerations into account.

\subsection{Comparison with state-of-the-art solutions}\label{sec:sota-comp}

\begin{table}[t!] 
	\begin{center}
		\begin{tabular}{ccc}
			\toprule
			{\bf Model} & {\bf Training time [min]} & {\bf No. of parameters} \\
			\midrule
			TCPCN (Ours) & $13$ & $153,711$  \\
			PN + GRU & $19$ & $218,115$  \\ 
			mm-GaitNet \cite{meng2020gait}& $32$ & $178,595$  \\ 
			bi-LSTM \cite{zhao2019mid}& $63$ & $3,237,379$  \\
			\bottomrule
		\end{tabular} 		
	\end{center}
	\caption{Comparison between TCPCN and other models from the literature in terms of training time and number of parameters. \label{tab:size-comp}}
\end{table}

Out of the two other approaches from the literature (see \secref{sec:rel-work}), \cite{meng2020gait}, does not obtain good results when subjects move freely, as neither a robust tracking method is implemented nor the identification information is used to improve the tracking performance,
while~\cite{zhao2019mid} performs the identification in an \textit{offline} fashion. In addition, they use different datasets. For these reasons, we chose to implement the classifiers from~\cite{meng2020gait} and~\cite{zhao2019mid} and evaluate them on our \mbox{multi-target} test dataset using $K=30$ input time steps and the same training data. As a baseline, we consider a model similar to TCPCN, but using a recurrent neural network (RNN) instead of temporal convolutions after the \mbox{point-cloud} feature extraction block. We refer to this model as \mbox{PN + GRU} in the following, as it is obtained combining a feature extraction block similar to PointNet with a \textit{gated recurrent unit} layer (GRU)~\cite{cho2014learning}, which is capable of learning \mbox{long-term} dependencies. GRU cells maintain a hidden state across time, processing it together with the current input vector to learn temporal features in the input sequence (see~\cite{cho2014learning} for a detailed description of GRU cells). In our implementation, we use a GRU layer with $128$ hidden units. 

In \tab{tab:size-comp}, we compare the learning models in terms of training time and number of parameters. This evaluation has been conducted on an NVIDIA RTX 2080 GPU for all the models. The training time is affected by the processing speed of each NN model and by the convergence time of the training process (number of training epochs). We note that the processing time of convolutional models (TCPCN and mm-GaitNet) is lower than that of recurrent ones (PN + GRU and bi-LSTM). However, training is significantly faster for the two models featuring the proposed point-cloud feature extractor (TCPCN and PN + GRU) due to faster convergence.

\begin{figure}[t!]
	\begin{center}   
		\includegraphics[width=8cm]{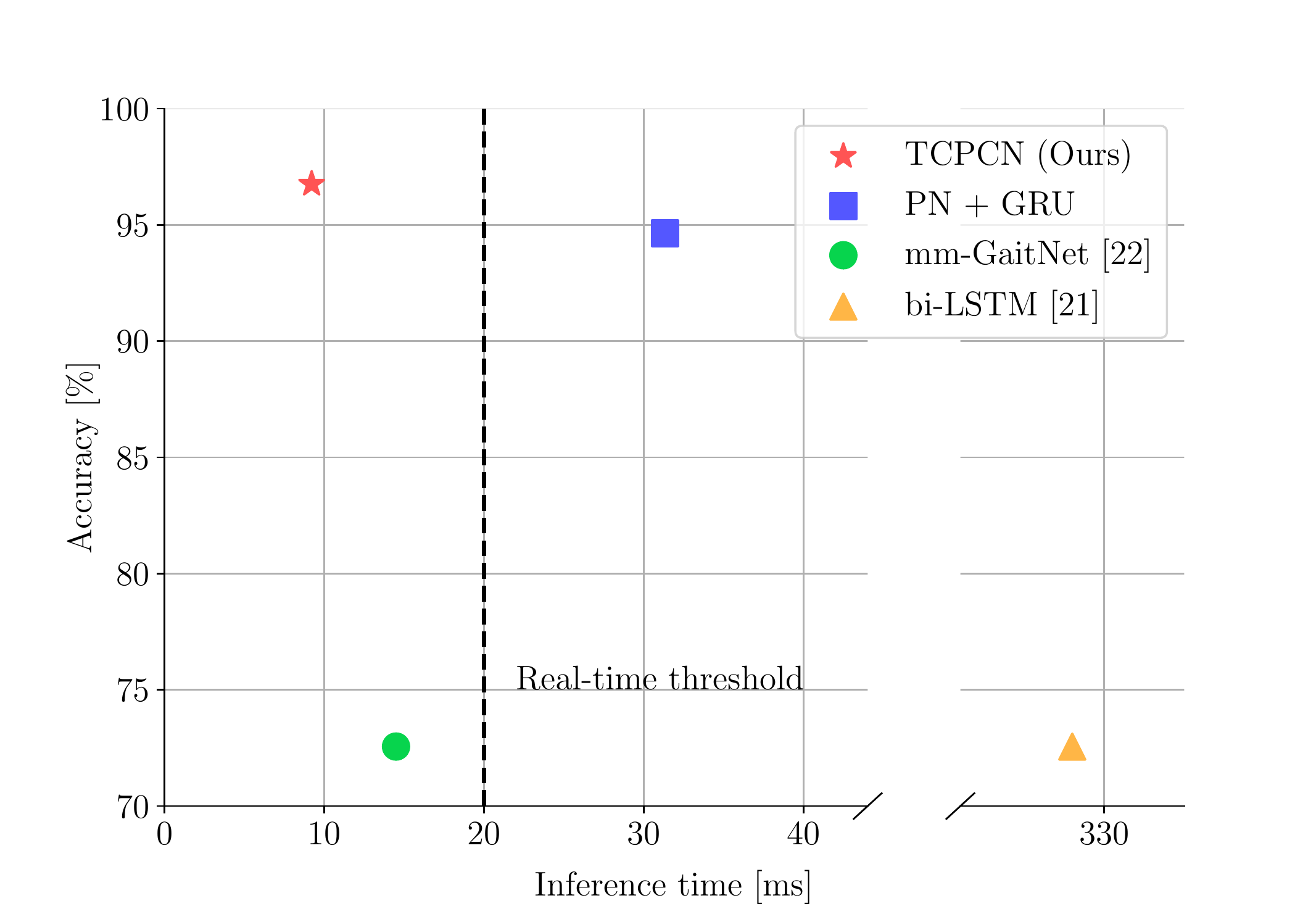} 
		\caption{Performance comparison of the proposed TCPCN model against mm-GaitNet~\cite{meng2020gait} and the bidirectional LSTM from~\cite{zhao2019mid}. As a baseline, we also evaluate a network similar to TCPCN that uses a GRU layer (PN + GRU) instead of temporal convolutions.}
		\label{fig:time-acc}
	\end{center}
\end{figure}

A comparison of accuracy and inference time, measured on the NVIDIA Jetson board, is presented in \fig{fig:time-acc}. The most accurate models in identifying the subjects are our TCPCN and PN + GRU. This shows the superiority of using a \mbox{point-cloud} feature extractor, due to its invariance to the ordering of the input points. TCPCN proves to be slightly better than PN + GRU, meaning that dilated temporal convolutions do not only improve the inference and training times but are also more effective in extracting temporal features. Through a vertical dashed line, we mark the maximum inference time for the algorithms to run in \mbox{real-time} on the Jetson device, i.e., $20$~ms (see \secref{sec:real-time}): only two models satisfy this constraint, namely the proposed TCPCN and \mbox{mm-GaitNet}~\cite{meng2020gait}, which both exploit convolutions, as opposed to the \mbox{RNN-based} \mbox{PN + GRU} and \mbox{bi-LSTM}. In particular, TCPCN is the fastest model in making predictions, with an average inference time of $9.21 \pm 2.12$~ms. 

%

\section{Concluding remarks}\label{sec:conclusion}

In this work, we proposed a novel system that performs real-time person tracking and identification on an edge computing device using sparse \mbox{point-cloud} data obtained from a \mbox{low-cost} \mbox{mm-wave} radar sensor. The raw signal undergoes several processing steps, including detection, clustering and Kalman filtering for position and subject extension estimation in the $x-y$ plane, followed by a fast neural network classifier based on a \mbox{point-cloud} specific feature extractor and dilated temporal convolutions. Our system significantly outperforms previous solutions from the literature, both in terms of accuracy and inference time, being able to reliably run in \mbox{real-time} at $15$~fps on an NVIDIA Jetson TX2 board, identifying up to three subject among a group of eight with an accuracy of almost $92 \%$, while simultaneously moving in an unseen indoor environment. 

Future research directions include the extension of the system to multiple radar devices, to deal with the frequent blockage events that can happen at \mbox{mm-wave} frequencies when multiple subjects move in the same physical environment. This would allow covering bigger spaces, while also getting better results in the presence of occlusions.

\bibliography{biblio}
\bibliographystyle{unsrt}

\vskip -20pt plus -1fil

\begin{IEEEbiography}[{\includegraphics[width=1in,height=1.25in,clip,keepaspectratio]{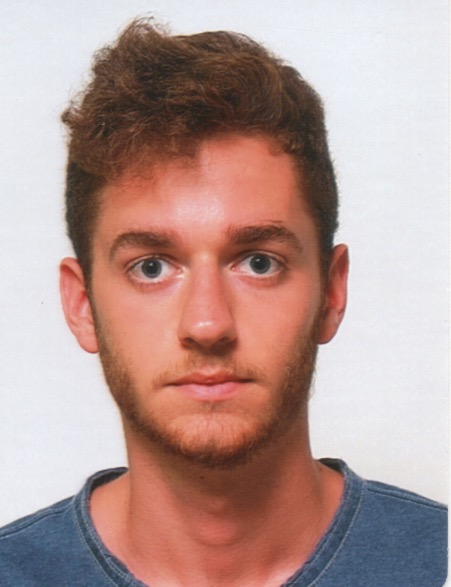}}]{Jacopo Pegoraro} (S'20) received the B.Sc. degree in information engineering and the M.Sc. degree in ICT for Internet and Multimedia engineering from the University of Padova, Padua, Italy, in 2017 and 2019, respectively. He is currently pursuing the Ph.D. degree with the SIGNET Research Group, Department of Information Engineering, in the same University. His research interests include deep learning and signal processing with applications to radio frequency sensing and, specifically, \mbox{mm-wave} radar sensing.
\end{IEEEbiography}

\vskip -20pt plus -1fil

\begin{IEEEbiography}[{\includegraphics[width=1in,height=1.25in,clip,keepaspectratio]{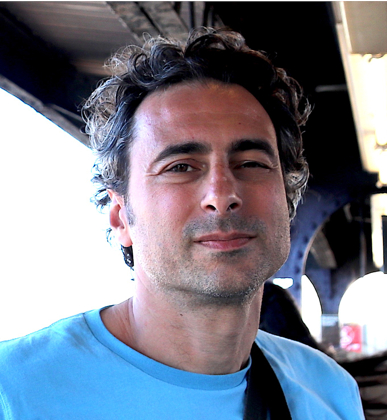}}]{Michele Rossi} (SM'13) is a Full Professor of Telecommunications in the Department of Information Engineering (DEI) at the University of Padova (UNIPD), Italy, teaching courses within the Master's Degree in ICT for internet and Multimedia (\url{http://mime.dei.unipd.it/}). He also sits on the Directive Board of the Master's Degree in Data Science offered by the Department of Mathematics (DM) at UNIPD (\url{https://datascience.math.unipd.it/}), for which he teaches machine learning and neural networks targeting the analysis of human data. Since 2017, he has been the Director of the DEI/IEEE Summer School of Information Engineering (\url{http://ssie.dei.unipd.it/}). His research interests lie in wireless sensing systems, green mobile networks, edge and wearable computing. In recent years, he has been involved in several EU projects on IoT technology (e.g., \mbox{IOT-A}, project no. 257521), and has collaborated with companies such as \mbox{DOCOMO} (compressive dissemination and network coding) and Worldsensing (IoT solutions for smart cities). In 2014, he has been the recipient of a SAMSUNG GRO award with a project entitled ``Boosting Efficiency in Biometric Signal Processing for Smart Wearable Devices''. In 2016-2018, he has been involved in the design of IoT protocols exploiting cognition and machine learning, as part of INTEL's Strategic Research Alliance (ISRA) R\&D program. His research is supported by the European Commission through several H2020 projects: SCAVENGE (no. 675891) on ``green 5G networks'', MINTS (no. 861222) on ``mm-wave networking and sensing'' and GREENEDGE (no. 953775) on ``green edge computing for mobile networks'' (project coordinator). Dr. Rossi has been the recipient of seven best paper awards from the IEEE and currently serves on the Editorial Boards of the IEEE Transactions on Mobile Computing, and of the Open Journal of the Communications Society.
\end{IEEEbiography}

\EOD

\end{document}